\documentclass[11pt]{article}

\usepackage{amssymb}
\usepackage{makeidx}
\usepackage[totalwidth=460truept,totalheight=600truept]{geometry}
\usepackage{latexsym,amssymb,amsmath,graphicx,accents,eucal,slashed,subfigure}
\usepackage{dsfont}
\usepackage[T1]{fontenc}
\usepackage{hyperref}

\def\theequation{\arabic{section}.\arabic{equation}}

\renewcommand{\theequation}{\thesection.\arabic{equation}}
\global\arraycolsep=1truept

\numberwithin{equation}{section}
\renewcommand{\theequation}{\arabic{section}.\arabic{equation}}

\begin{document}

\bigskip \hfill IFUP-TH/2014-2

\phantom{C}

\vskip 1.4truecm

\begin{center}
{\huge \textbf{Adler-Bardeen Theorem}}

\vskip .5truecm

{\huge \textbf{And Manifest Anomaly Cancellation}}

\vskip .5truecm

{\huge \textbf{To All Orders In Gauge Theories}}

\vskip 1truecm

\textsl{Damiano Anselmi}

\vskip .2truecm

\textit{Dipartimento di Fisica ``Enrico Fermi'', Universit\`{a} di Pisa, }

\textit{and INFN, Sezione di Pisa,}

\textit{Largo B. Pontecorvo 3, I-56127 Pisa, Italy}

\vskip .2truecm

damiano.anselmi@df.unipi.it

\vskip 1.5truecm

\textbf{Abstract}
\end{center}

\medskip

We reconsider the Adler-Bardeen theorem for the cancellation of gauge
anomalies to all orders, when they vanish at one loop. Using the
Batalin-Vilkovisky formalism and combining the dimensional-regularization
technique with the higher-derivative gauge invariant regularization, we
prove the theorem in the most general perturbatively unitary renormalizable
gauge theories coupled to matter in four dimensions, and identify the
subtraction scheme where anomaly cancellation to all orders is manifest,
namely no subtractions of finite local counterterms are required from two
loops onwards. Our approach is based on an order-by-order analysis of
renormalization, and, differently from most derivations existing in the
literature, does not make use of arguments based on the properties of the
renormalization group. As a consequence, the proof we give also applies to
conformal field theories and finite theories.

\vskip 1truecm

\vfill\eject

\section{Introduction}

\label{RGren}\setcounter{equation}{0}

The Adler-Bardeen theorem \cite{adlerbardeen,review} is a crucial property
of quantum field theory, and one of the few tools to derive exact results.
In the literature various statements go under the name of \textquotedblleft
Adler-Bardeen theorem\textquotedblright . They apply to different
situations. The original statement by Adler and Bardeen says that (I) the
Adler-Bell-Jackiw axial anomaly \cite{ABJ} is one-loop exact. The second
statement, which is the one we are going to study here, says that (II)
(there exists a subtraction scheme where) gauge anomalies vanish to all
orders, if they vanish at one loop. Statement II is important to justify the
cancellation of gauge anomalies to all orders in the standard model. A third
statement concerns the one-loop exactness of anomalies associated with
external fields.

Statement I is expressed by a well-known operator identity for the
divergence of the axial current. By means of a diagrammatic analysis, Adler
and Bardeen were able to provide the subtraction scheme where that identity
is \textit{manifestly} one-loop exact in QED \cite{adlerbardeen}. They
emphasized that higher-order corrections vanish, unless they contain the
one-loop triangle diagram as a subdiagram. Said like this, statement I
intuitively implies statement II. However, the original proof of Adler and
Bardeen applies only to QED.

Other approaches to the problem have appeared, since the paper by Adler and
Bardeen, in Abelian and non-Abelian gauge theories \cite{review}. Statement
I can be proved using arguments based on the properties of the
renormalization group \cite{zee,collins,tonin}, regularization independent
algebraic techniques \cite{lucchesi}, or an algebraic/geometric derivation 
\cite{witten} based on the Wess-Zumino consistency conditions \cite%
{wesszumino} and the quantization of the Wess-Zumino-Witten action.
Statement II can also be proved using renormalization-group (RG)\ arguments,
with the dimensional regularization \cite{tonin2} or
regularization-independent approaches \cite{sorella}.

More recently, statement II was proved by the author of this paper in
standard model extensions with high-energy Lorentz violation \cite{lvsm},
which are renormalizable by \textquotedblleft weighted power
counting\textquotedblright\ \cite{halat}. The approach of \cite{lvsm} is
closer to the original approach by Adler and Bardeen, in the sense that it
does not make use of RG arguments, algebraic methods or geometric shortcuts,
it naturally provides the subtraction scheme where the all-order
cancellation is manifest, and it is basically a diagrammatic analysis,
although instead of dealing directly with diagrams, it uses the
Batalin-Vilkovisky formalism \cite{bata} to manage relations among diagrams
in a compact and efficient way.

In the present paper we prove statement II\ in the most general
perturbatively unitary, renormalizable gauge theories coupled to matter, and
elaborate further along the guidelines of ref. \cite{lvsm}. We upgrade the
approach of \cite{lvsm} in a number of directions, emphasize properties that
were not apparent at that time, and expand the arguments that were presented
concisely. We also gain a certain clarity by dropping the Lorentz violation.
A side purpose of this investigation is to develop new techniques and tools
to prove all-order theorems in quantum field theory with a smaller effort.

Our results make progress in several directions. To our knowledge, if we
exclude ref. \cite{lvsm} and this paper, statement II has been proved beyond
QED only making use of arguments based on the renormalization group.
However, RG arguments do not provide the subtraction scheme where the
all-order cancellation is manifest, and are not sufficiently general. For
example, they are powerless when the beta functions identically vanish, so
they exclude conformal field theories and finite theories, where however the
Adler-Bardeen theorem does hold. Actually, RG techniques fail even when the
first coefficients of the beta functions vanish \cite{tonin2,sorella}. Our
approach does not suffer from these limitations. Another reason to avoid
shortcuts is that in the past the Adler-Bardeen theorem caused some
confusion in the literature, therefore new proofs, and even more
generalizations, should be as transparent as possible. In this paper we pay
attention to all details.

The all-order cancellation of gauge anomalies is a property that depends on
the scheme, but the existence of a good scheme is not evident. Knowing the
scheme where the cancellation is manifest is very convenient from the
practical point of view, because it saves the effort of subtracting \textit{%
ad hoc} finite local counterterms at each step of the perturbative
expansion. For example, using the dimensional regularization and the minimal
subtraction scheme the cancellation of two-loop and higher-order corrections
to gauge anomalies in the standard model is not manifest, and finite local
counterterms must be subtracted every time.

To find the right subtraction scheme we need to define a clever
regularization technique. It turns out that using the Batalin-Vilkovisky
formalism and combining the dimensional regularization with the gauge
invariant higher-derivative regularization, the subtraction scheme where the
Adler-Bardeen theorem is manifest emerges quite naturally \cite{lvsm}.

It is well-known that, in general, gauge invariant higher-derivative
regularizations do not regularize completely, because some one-loop diagrams
can remain divergent. From our viewpoint, this is not a weakness, because it
allows us to separate the sources of potential anomalies from everything
else. We just have to use a second regulator, the dimensional one, to deal
with the few surviving divergent diagrams.

The regularization we are going to use introduces two cutoffs:\ $\varepsilon
=4-D$, where $D$ is the continued complex dimension, and an energy scale $%
\Lambda $ for the higher-derivative regularizing terms. The regularized
action must be gauge invariant in $D=4$, to ensure that the
higher-derivative regulator has the minimum impact on gauge anomalies. The
physical limit is defined letting $\varepsilon $ tend to $0$ and $\Lambda $
to $\infty $. When we have two or more cutoffs, physical quantities do not
depend on the order in which we remove them. More precisely, exchanging the
order of the limits $\varepsilon \rightarrow 0$ and $\Lambda \rightarrow
\infty $ is equivalent to change the subtraction scheme. That kind of scheme
change is however crucial for our arguments.

Consider first the limit $\Lambda \rightarrow \infty $ followed by $%
\varepsilon \rightarrow 0$. When $D\neq 4$ the limit $\Lambda \rightarrow
\infty $ is regular in every diagram and gives back the dimensionally
regularized theory: no $\Lambda $ divergences appear, but just poles in $%
\varepsilon $. In this framework there are no known subtraction schemes
where the Adler-Bardeen theorem holds manifestly.

Now, consider the limit $\varepsilon \rightarrow 0$ followed by $\Lambda
\rightarrow \infty $. At fixed $\Lambda $ we have a higher-derivative
theory. If properly organized, that theory is superrenormalizable and
contains just a few (one-loop) divergent diagrams, which are poles in $%
\varepsilon $ and may be removed by redefining some parameters. At a second
stage, we study the limit $\Lambda \rightarrow \infty $, where $\Lambda $
divergences appear and are removed by redefining parameters and making
canonical transformations. We call the regularization technique defined this
way \textit{dimensional/higher-derivative} (DHD) regularization.

Intuitively, if gauge anomalies are trivial at one loop, there should be no
further problems at higher orders, because the higher-derivative
regularization is manifestly gauge invariant. Thus, we expect that the DHD
regularization provides the framework where the Adler-Bardeen theorem is
manifest. However, it is not entirely obvious that the two regularization
techniques can be merged to achieve the goal we want. Among the other
things, $\varepsilon $ evanescent terms are around all the time and the $%
\mathcal{O}(1/\Lambda ^{n})$ regularizing terms can simplify power-like $%
\Lambda $ divergences, causing troubles. Nevertheless, with some effort and
a nontrivial amount of work we can prove that all difficulties can be
properly dealt with.

Summarizing, the statement we prove in this paper is

\textbf{Theorem}. \textit{In renormalizable perturbatively unitary gauge
theories coupled to matter, there exists a subtraction scheme where gauge
anomalies manifestly cancel to all orders, if they are trivial at one loop.%
\label{bardo}}

Once we have this result, we know that no matter what scheme we use, it is
always possible to find \textit{ad hoc} finite local counterterms that
ensure the cancellation of gauge anomalies to all orders. Then we are free
to use the more common minimal subtraction scheme and the pure dimensional
regularization technique.

The paper is organized as follows. In sections 2-7 we prove the theorem in
non-Abelian Yang-Mills theory coupled to left-handed chiral fermions. This
model is sufficiently general to illustrate the key points of the proof, as
well as the main arguments and tools, but relatively simple to free the
derivation from unnecessary complications. At the end of the paper, in
section 8, we show how to include the missing fields, namely right-handed
fermions, scalars and photons, and cover the most general perturbatively
unitary renormalizable gauge theory coupled to matter. Section 9 contains
our conclusions. In appendix A we recall the calculation of gauge anomalies
in chiral theories. In appendix B we recall the proof of a useful formula.

The proof for Yang-Mills theory coupled to chiral fermions is organized as
follows. In sections 2 and 3 we formulate the dimensional and DHD
regularization techniques. In sections 4-6 we prove the Adler-Bardeen
theorem in the higher-derivative theory, studying the limit $\varepsilon
\rightarrow 0$ at $\Lambda $ fixed. Precisely, in section 4 we work out the
renormalization, in section 5 we study the one-loop anomalies and in section
6 we prove the anomaly cancellation to all orders. In section 7 we take the
limit $\Lambda \rightarrow \infty $ and conclude the proof of the
Adler-Bardeen theorem for the final theory.

\section{Dimensional regularization of chiral Yang-Mills theory}

\setcounter{equation}{0}

We first prove the Adler-Bardeen theorem in detail in four-dimensional
non-Abelian Yang-Mills theory coupled to left-handed chiral fermions. This
model offers a sufficiently general arena to illustrate the key arguments
and tools of our approach. At the same time, we make some clever choices to
prepare the generalization (discussed in section 8) to the most general
perturbatively unitary gauge theories coupled to matter. To begin with, in
this section we dimensionally regularize chiral gauge theories and point out
a number of facts and properties that are normally not emphasized, but are
rather important for the arguments of this paper.

Consider a gauge theory with gauge group $G$ and left-handed chiral fermions 
$\psi _{L}^{I}$ in certain irreducible representations $R_{L}^{I}$ of $G$.
If $G$ is the product of various simple groups $G_{i}$, we use indices $%
a,b,\ldots $ for $G$ and indices $a_{i},b_{i},\ldots $ for $G_{i}$. Denote
the gauge coupling $g_{i}$ of each $G_{i}$ with $gr_{i}$, where $r_{i}$ are
parameters of order one that we incorporate into the $G$ structure constants 
$f^{abc}$ and the anti-Hermitian matrices $T^{a}$ associated with the
representations of matter fields. We call $g$ the \textit{overall gauge
coupling}. We organize the matrices $T^{a}$ in block-diagonal form, where
each block refers to a $\psi _{L}^{I}$ and its representation $R_{L}^{I}$.
When we write $T^{a}\psi _{L}^{I}$ we understand that $T^{a}$ is replaced by
the appropriate block. More fermions in the same irreducible representations
may be present. With these conventions, the matrices $T^{a}$ still satisfy $%
[T^{a},T^{b}]=f^{abc}T^{c}$ and the classical action reads 
\begin{equation}
S_{c}=-\frac{1}{4}\sum_{i}\zeta _{i}\int F_{\mu \nu }^{a_{i}}F^{a_{i}\hspace{%
0.01in}\mu \nu }+\int \bar{\psi}_{L}^{I}\imath \slashed{D}\psi _{L}^{I},
\label{elle}
\end{equation}%
where $F_{\mu \nu }^{a_{i}}=\partial _{\mu }A_{\nu }^{a_{i}}-\partial _{\nu
}A_{\mu }^{a_{i}}+g_{i}f^{a_{i}b_{i}c_{i}}A_{\mu }^{b_{i}}A_{\nu }^{c_{i}}$
(no sum over this kind of index $i$ being understood, here and in the rest
of the paper) is the $G_{i}$ field strength, $D_{\mu }\psi _{L}^{I}=\partial
_{\mu }\psi _{L}^{I}+gT^{a}A_{\mu }^{a}\psi _{L}^{I}$ is the fermion
covariant derivative and $\imath $ is used for $\sqrt{-1}$ to avoid
confusion with the index $i$. The parameters $\zeta _{i}$ could be
normalized to 1, but for future uses it is convenient to keep them free,
because they are renormalized by poles in $\varepsilon $. Analogous
parameters in front of the fermionic kinetic terms are not necessary.

To keep the presentation simple we make some simplifying assumptions that do
not restrict the validity of our arguments. Specifically, we do not include
right-handed fermions and scalar fields, and assume that the groups $G_{i}$
are non-Abelian, so there is no renormalization mixing among gauge fields,
even when more copies of the same simple group are present. In section 8 we
explain how to relax these assumptions and cover the most general Abelian
and non-Abelian perturbatively unitary renormalizable gauge theories coupled
to matter.

Let us briefly recall the Batalin-Vilkovisky formalism for general gauge
theories \cite{bata}. The classical fields $\phi =\{A_{\mu }^{a},\psi
_{L}^{I},\bar{\psi}_{L}^{I}\}$, together with the ghosts $C$, the antighosts 
$\bar{C}$ and the Lagrange multipliers $B$ for the gauge fixing are
collected into the set of fields $\Phi ^{\alpha }=\{A_{\mu }^{a},C^{a},\bar{C%
}^{a},B^{a},\psi _{L}^{I},\bar{\psi}_{L}^{I}\}$. An external source $%
K_{\alpha }$ with opposite statistics is associated with each $\Phi ^{\alpha
}$, and coupled to the $\Phi ^{\alpha }$ transformations $R^{\alpha }(\Phi
,g)$. We have $K_{\alpha }=\{K^{\mu a},K_{C}^{a},K_{\bar{C}%
}^{a},K_{B}^{a},K_{\psi }^{I},\bar{K}_{\psi }^{I}\}$. If $X$ and $Y$ are
functionals of $\Phi $ and $K$ their \textit{antiparentheses} are defined as 
\begin{equation}
(X,Y)\equiv \int \left( \frac{\delta _{r}X}{\delta \Phi ^{\alpha }}\frac{%
\delta _{l}Y}{\delta K_{\alpha }}-\frac{\delta _{r}X}{\delta K_{\alpha }}%
\frac{\delta _{l}Y}{\delta \Phi ^{\alpha }}\right) ,  \label{usa}
\end{equation}
where the integral is over spacetime points associated with repeated
indices. The \textit{master equation} $(S,S)=0$ must be solved with the
``boundary condition'' $S(\Phi ,K)=S_{c}(\phi )$ at $C=\bar{C}=B=K=0$ in $%
D=4 $, where $S_{c}(\phi )$ is the classical action (\ref{elle}). The
solution $S(\Phi ,K)$ is the action we start with to quantize the theory.

In the model we are considering the gauge algebra closes off shell, so there
exists a variable frame where $S(\Phi ,K)$ is linear in $K$. The
non-gauge-fixed solution of the master equation is 
\begin{equation*}
S_{\text{ngf}}(\Phi ,K)=S_{c}(\phi )+S_{K},
\end{equation*}%
where the functional%
\begin{eqnarray*}
S_{K}(\Phi ,K) &=&-\int R^{\alpha }(\Phi ,g)K_{\alpha }=-\int (D_{\mu
}C^{a})K^{\mu a}+\frac{g}{2}\int f^{abc}C^{b}C^{c}K_{C}^{a}-\int B^{a}K_{%
\bar{C}}^{a} \\
&&+g\int \left( \bar{\psi}_{L}^{I}T^{a}C^{a}K_{\psi }^{I}+\bar{K}_{\psi
}^{I}T^{a}C^{a}\psi _{L}^{I}\right)
\end{eqnarray*}%
collects the symmetry transformations of the fields, $D_{\mu }C^{a}=\partial
_{\mu }C^{a}+gf^{abc}A_{\mu }^{b}C^{c}$ being the covariant derivative of
the ghosts. The gauge-fixed solution of the master equation reads 
\begin{equation}
S_{\text{gf}}(\Phi ,K)=S_{\text{ngf}}+(S_{K},\Psi )=S_{c}(\phi )+(S_{K},\Psi
)+S_{K},  \label{snev}
\end{equation}%
where $\Psi (\Phi )$ is the \textquotedblleft gauge
fermion\textquotedblright , a functional of ghost number $-1$ that collects
the gauge-fixing conditions. For convenience, we choose standard linear
gauge-fixing conditions and write 
\begin{equation}
\Psi (\Phi )=\int \sum_{i}\bar{C}^{a_{i}}\left( \partial ^{\mu }A_{\mu
}^{a_{i}}+\frac{\xi _{i}}{2}B^{a_{i}}\right)  \label{gf}
\end{equation}%
where $\xi _{i}$ are gauge-fixing parameters.

The na\"{\i}ve $D$-dimensional continuation of the action (\ref{elle}) is
not well regularized, because chiral fermions do not have good propagators.
To overcome this difficulty, we proceed as follows. As usual, we split the $%
D $-dimensional spacetime manifold $\mathbb{R}^{D}$ into the product $%
\mathbb{R}^{4}\times \mathbb{R}^{-\varepsilon }$ of ordinary
four-dimensional spacetime $\mathbb{R}^{4}$ times a residual $(-\varepsilon
) $-dimensional evanescent space $\mathbb{R}^{-\varepsilon }$. Spacetime
indices $\mu ,\nu ,\ldots $ of vectors and tensors are split into bar
indices $\bar{\mu},\bar{\nu},\ldots $, which take the values 0,1,2,3, and
formal hat indices $\hat{\mu},\hat{\nu},\ldots $, which denote the $\mathbb{R%
}^{-\varepsilon }$ components. For example, momenta $p^{\mu }$ are split
into pairs $p^{\bar{\mu}}$, $p^{\hat{\mu}}$, or equivalently $\bar{p}^{\mu }$%
, $\hat{p}^{\mu }$. The flat-space metric $\eta _{\mu \nu }=$diag$%
(1,-1,\ldots ,-1)$ is split into $\eta _{\bar{\mu}\bar{\nu}}=$diag$%
(1,-1,-1,-1)$ and $\eta _{\hat{\mu}\hat{\nu}}=-\delta _{\hat{\mu}\hat{\nu}}$%
. When we contract evanescent components we use the metric $\eta _{\hat{\mu}%
\hat{\nu}}$, so for example $\hat{p}^{2}=p^{\hat{\mu}}\eta _{\hat{\mu}\hat{%
\nu}}p^{\hat{\nu}}$. We assume that the continued $\gamma $ matrices $\gamma
^{\mu }$ satisfy the continued Dirac algebra $\{\gamma ^{\mu },\gamma ^{\nu
}\}=2\eta ^{\mu \nu }$. We define $\gamma _{5}=\imath \gamma ^{0}\gamma
^{1}\gamma ^{2}\gamma ^{3}$, $P_{L}=(1-\gamma _{5})/2$, $P_{R}=(1+\gamma
_{5})/2$ and the charge-conjugation matrix $C=-\imath \gamma ^{0}\gamma ^{2}$
in the usual fashion. Full $SO(1,D-1)$ invariance is lost in most
expressions, replaced $SO(1,3)\times SO(-\varepsilon )$ invariance.

The action (\ref{elle}) gives the fermion propagator $P_{L}(\imath /%
\slashed{\bar{p}})P_{R}$, which involves only the four-dimensional
components $\bar{p}^{\mu }$ of momenta. Therefore, it does not fall off in
all directions of integration for $p\rightarrow \infty $. Applying the rules
of the dimensional regularization, fermion loops integrate to zero. To
provide fermions with correct propagators we introduce right-handed $\psi
_{L}^{I}$-partners $\psi _{R}^{I}$ that decouple in four dimensions and are
inert under every gauge transformations. We include $\psi _{R}$ and $\bar{%
\psi}_{R}$ into the set of fields $\Phi $. It is not necessary to introduce
sources $K$ for them.

Specifically, we start from the regularized classical action 
\begin{equation}
S_{c\text{r}}=-\frac{1}{4}\sum_{i}\zeta _{i}\int F_{\mu \nu }^{a_{i}}F^{a_{i}%
\hspace{0.01in}\mu \nu }+\int \bar{\psi}_{L}^{I}\imath \slashed{D}\psi
_{L}^{I}+S_{\text{LR}}=S_{c}+S_{\text{LR}},  \label{buh2}
\end{equation}%
which is the sum of the unregularized classical action (\ref{elle}) plus a
correction 
\begin{equation}
S_{\text{LR}}=\varsigma _{IJ}\int \bar{\psi}_{R}^{I}\imath \slashed{\partial}%
\psi _{L}^{J}+\varsigma _{JI}^{\ast }\int \bar{\psi}_{L}^{I}\imath %
\slashed{\partial}\psi _{R}^{J}+\int \bar{\psi}_{R}^{I}\imath %
\slashed{\partial}\psi _{R}^{I},  \label{slr}
\end{equation}%
where $\varsigma _{IJ}$ are constants that form an invertible matrix $%
\varsigma $. The only nontrivial off-diagonal entries of $\varsigma $ (and
of all the matrices $M_{IJ}$ we going to meet in this paper) are those that
mix equivalent irreducible representations $R_{L}^{I}$. The reason why the
matrix $\varsigma $ is kept free is that later on it will help us reabsorb
the renormalization constants of $\psi _{L}^{I}$, since $S_{\text{LR}}$ is
nonrenormalized (see below).

Using the polar decomposition, we can write $\varsigma =U_{R}^{\dagger
}DU_{L}$, where $U_{L}$ and $U_{R}$ are unitary matrices and $D$ is a
positive-definite diagonal matrix. In the basis where $\varsigma $ is
replaced by its diagonal form $D\equiv $ diag$(\varsigma _{I})$ the
propagators of the Dirac fermions $\psi ^{I}=\psi _{L}^{I}+\psi _{R}^{I}$
are 
\begin{equation}
\imath \delta ^{IJ}\frac{\slashed{\bar{p}}+\varsigma _{I}\slashed{\hat{p}}}{%
\bar{p}^{2}+\varsigma _{I}^{2}\hat{p}^{2}}  \label{fpar}
\end{equation}
and coincide with the usual propagators for $\varsigma _{I}=1$.

Next, observe that $(S_{K},S_{K})=0$ in arbitrary $D$. The regularized
gauge-fixed action is (up to an extension that will be discussed later) 
\begin{equation}
S_{r0}(\Phi ,K)=S_{c}+S_{\text{LR}}+(S_{K},\Psi )+S_{K}=S_{\text{gf}}+S_{%
\text{LR}},  \label{sr}
\end{equation}
and satisfies 
\begin{equation}
(S_{r0},S_{r0})=2\imath g\int C^{a}\left( (\partial _{\hat{\mu}}\bar{\psi}%
_{R}^{I})\gamma ^{\hat{\mu}}T^{a}\varsigma _{IJ}\psi _{L}^{J}+\bar{\psi}%
_{L}^{I}\varsigma _{JI}^{*}T^{a}\hat{\slashed{\partial}}\psi _{R}^{J}\right)
=\mathcal{O}(\varepsilon ),  \label{srsr}
\end{equation}
where ``$\mathcal{O}(\varepsilon )$'' is used to denote any expression that
vanishes in four dimensions. We have used $P_{R}\slashed{\partial}P_{R}=P_{R}%
\hat{\slashed{\partial}}P_{R}$ and a similar relation with $R\rightarrow L$.
Observe that $S_{r0}$ is invariant under the global symmetry transformations
of the group $G$.

Given a (dimensionally) regularized classical action $S(\Phi ,K)$, the
regularized generating functionals $Z$ and $W$ are defined by the formulas 
\begin{equation}
Z(J,K)=\int [\mathrm{d}\Phi ]\exp \left( \imath S(\Phi ,K)+\imath \int \Phi
^{\alpha }J_{\alpha }\right) =\exp \imath W(J,K),  \label{zg}
\end{equation}%
and the generating functional $\Gamma (\Phi ,K)=W(J,K)-\int \Phi ^{\alpha
}J_{\alpha }$ of one-particle irreducible diagrams is the Legendre transform
of $W(J,K)$ with respect to $J$, where the sources $K$ act as spectators.
Often it is necessary to pay attention to the action used to define
averages. We denote the averages $\langle \cdots \rangle $ defined by the
action $S$ as $\langle \cdots \rangle _{S}$ (at $J_{\alpha }\neq 0$). The
anomaly functional is 
\begin{equation}
\mathcal{A}=(\Gamma ,\Gamma )=\langle (S,S)\rangle _{S}  \label{anom}
\end{equation}%
and collects the set of one-particle irreducible correlation functions
containing one insertion of $(S,S)$. The last equality of formula (\ref{anom}%
) can be proved by making the change of field variables $\Phi ^{\alpha
}\rightarrow \Phi ^{\alpha }+\theta (S,\Phi ^{\alpha })$ inside the
functional integral (\ref{zg}), where $\theta $ is a constant anticommuting
parameter. The proof is recalled in appendix B, together with comments on
the meaning of the formula.

No one-particle irreducible diagrams can be constructed with external legs $%
\bar{\psi}_{R}$ or $\psi _{R}$, because $\bar{\psi}_{R}$ and $\psi _{R}$ do
not appear in any vertices. Thus, the total $\Gamma $ functional satisfies 
\begin{equation*}
\Gamma (\Phi ,K)=\left. \Gamma (\Phi ,K)\right| _{\bar{\psi}_{R}=\psi
_{R}=0}+S_{\text{LR}}.
\end{equation*}

We have anticipated that the action (\ref{sr}) is not the final
dimensionally regularized action we are going to use. Before moving to the
appropriate extension $S_{r}$, we must describe the counterterms generated
by $S_{r0}$, list a number of properties that can be used to restrict the $%
S_{r0}$ extensions and point out some subtleties concerning the dimensional
regularization.

First, observe that the counterterms are $B$, $K_{B}$ and $K_{\bar{C}}$
independent. Indeed, the source $K_{B}$ appears nowhere in $S_{r0}$, while $%
K_{\bar{C}}$ appears only in $-\int BK_{\bar{C}}$. Moreover, the gauge
fixing conditions are linear in the fields, and the $B$-dependent terms of $%
S_{r0}$ are at most quadratic in $\Phi $. Therefore, no nontrivial
one-particle irreducible diagrams can have external $B$ legs.

Second, the action $S_{r0}$ does not depend on the antighosts $\bar{C}%
^{a_{i}}$ and the sources $K^{\mu a_{i}}$ separately, but only through the
combinations $K^{\mu a_{i}}+\partial _{\mu }\bar{C}^{a_{i}}$. The $\Gamma $
functional must share the same property. Indeed, an antighost external leg
actually carries the structure $\partial _{\mu }\bar{C}^{a_{i}}$, since all
vertices containing antighosts do so. Given a diagram with $K^{\mu a_{i}}$
or $\partial _{\mu }\bar{C}^{a_{i}}$ on external legs, we can construct
almost identical diagrams by just replacing one or more legs $K^{\mu a_{i}}$
with $\partial _{\mu }\bar{C}^{a_{i}}$, or vice versa.

Third, power counting and ghost-number conservation ensure that the
counterterms are linear in the sources $K$. Using square brackets to denote
dimensions in units of mass, we have $[K^{\hspace{0.01in}\mu
a}]=[K_{C}^{a}]=2$, and $[K_{\psi }]=3/2$. These sources have negative ghost
numbers. Therefore, the dimension of a term that is more than linear in $K$
and has vanishing ghost number necessarily exceeds 4.

\subsection{Structure of the dependence on the overall gauge coupling}

It is useful to single out how the functionals depend on the overall gauge
coupling $g$. The tree-level functionals we work with have the $g$ structure 
\begin{equation}
X_{\text{tree}}(\Phi ,K,g)=\frac{1}{g^{2}}X_{\text{tree}}^{\prime }(g\Phi
,gK).  \label{liable}
\end{equation}%
If the action satisfies this condition at the tree level, then the
renormalized action and the $\Gamma $ functional have the $g$ structure 
\begin{equation}
X(\Phi ,K,g)=\sum_{L\geqslant 0}g^{2(L-1)}X_{L}^{\prime }(g\Phi ,gK),
\label{liable2}
\end{equation}%
where $X_{L}$ collects the $L$-loop contributions. Basically, there is an
additional factor $g^{2}$ for every loop. Indeed, when the action is of the
form (\ref{liable}), every vertex is multiplied by a power $g^{N-2}$, where $%
N$ is the number of its $\Phi $ plus $K$ legs. Then, a one-particle
irreducible diagram with $L$ loops, $I$ internal legs, $E$ external legs and 
$v_{i}$ vertices with $i$ legs is multiplied by 
\begin{equation*}
\prod_{i\geqslant 2}g^{v_{i}(i-2)}=g^{2I+E-2V}=g^{E-2}g^{2L}=g^{E}g^{2(L-1)},
\end{equation*}%
having used $L-I+V=1$ and $\sum_{i\geqslant 2}iv_{i}=2I+E$. We see that for $%
L\geq 1$ we have one power of $g$ for each external leg and a residual
factor $g^{2(L-1)}$, in agreement with (\ref{liable2}).

The $g$ structures (\ref{liable}) and (\ref{liable2})\ are preserved by the
antiparentheses: if the functionals $X(\Phi ,K,g)$ and $Y(\Phi ,K,g)$
satisfy (\ref{liable}), or (\ref{liable2}), then the functional $(X,Y)$
satisfies (\ref{liable}), or (\ref{liable2}), respectively.

\subsection{Properties of the dimensional regularization of chiral theories}

Now we recall a few properties of the dimensional regularization of chiral
theories, which are important for the rest of our analysis. It is well-known
that divergences are just poles in $\varepsilon $. Instead, the terms that
disappear when $D\rightarrow 4$, called ``evanescences'', can be of two
types: \textit{formal} or \textit{analytic}. Analytically evanescent terms,
briefly denoted by ``aev'', are those that factorize at least one $%
\varepsilon $, such as $\varepsilon F_{\mu \nu }F^{\mu \nu }$, $\varepsilon 
\bar{\psi}_{L}\imath \slashed{D}\psi _{L}$, etc. Formally evanescent terms,
briefly denoted by ``fev'', are those that formally disappear when $%
D\rightarrow 4$, but do not factorize powers of $\varepsilon $. They are
built with the tensor $\delta _{\hat{\mu}\hat{\nu}}$ and the evanescent
components $\hat{x}$, $\hat{p}$, $\hat{\partial}$, $\hat{\gamma}$, $\hat{A}$
of coordinates, momenta, derivatives, gamma matrices and gauge fields.
Examples are $\bar{\psi}_{L}\imath \hat{\slashed{\partial}}\psi _{R}$, $%
(\partial _{\hat{\mu}}A_{\nu }^{a})(\partial ^{\hat{\mu}}A^{\nu a})$, etc.

The distinction between formally evanescent and analytically evanescent
expressions is to some extent ambiguous. Consider for example a basis $\bar{%
\psi}_{1}\gamma ^{\rho _{1}\cdots \rho _{k}}\psi _{2}$ of fermion bilinears,
where $\psi _{1}$, $\psi _{2}$ can be $\psi _{L}$ or $K_{\psi }$, and $%
\gamma ^{\rho _{1}\cdots \rho _{k}}$ is the completely antisymmetric product
of $\gamma ^{\rho _{1}},\cdots ,\gamma ^{\rho _{k}}$. In dimensional
regularization these bilinears are nonvanishing for every $k$, and they are
evanescent for $k>4$. We have several ways to rearrange the products of two
or more fermion bilinears by using Fierz identities, and such rearrangements
can convert formally evanescent objects into analytically evanescent ones.
For example, given some spinors $\psi _{n}$, $n=1,2,3,4$, we can expand the
matrix $\psi _{2}\bar{\psi}_{3}$ in the basis made of $\gamma ^{\rho
_{1}\cdots \rho _{k}}$, $k=0,\ldots ,\infty $. We have 
\begin{equation*}
\psi _{2}\bar{\psi}_{3}=-\frac{1}{f(D)}\sum_{k=0}^{\infty }\frac{%
(-1)^{k(k-1)/2}}{k!}\gamma ^{\rho _{1}\cdots \rho _{k}}(\bar{\psi}_{3}\gamma
_{\rho _{1}\cdots \rho _{k}}\psi _{2}),
\end{equation*}%
where $f(D)=$tr$[\mathds{1}]$. Using this identity we find, for example, 
\begin{equation}
(\bar{\psi}_{1}\gamma ^{\hat{\mu}}\psi _{2})(\bar{\psi}_{3}\gamma _{\hat{\mu}%
}\psi _{4})=\frac{\varepsilon }{f(D)}(\bar{\psi}_{1}\psi _{4})(\bar{\psi}%
_{3}\psi _{2})-\frac{2}{f(D)}(\bar{\psi}_{1}\gamma ^{\hat{\rho}}\psi _{4})(%
\bar{\psi}_{3}\gamma _{\hat{\rho}}\psi _{2})-\frac{\varepsilon }{f(D)}(\bar{%
\psi}_{1}\gamma ^{\rho }\psi _{4})(\bar{\psi}_{3}\gamma _{\rho }\psi
_{2})+\cdots  \label{divam}
\end{equation}%
Basically, this equation has the form \textquotedblleft fev $=$ fev $+$
aev\textquotedblright . The existence of such relations poses some problems,
which we now describe.

Feynman diagrams may generate ``divergent evanescences'', briefly denoted by
``divev''. They are made of products between poles and formal evanescences,
such as $(\partial _{\hat{\mu}}A_{\nu }^{a})(\partial ^{\hat{\mu}}A^{\nu
a})/\varepsilon $. The theorem of locality of counterterms demands that we
renormalize divergent evanescences away, together with ordinary divergences
(see below). However, this makes sense only if we can define divergent
evanescences unambiguously, which could be problematic due to the
observations made above. For example, if we multiply both sides of formula (%
\ref{divam}) by $1/\varepsilon $ we get a relation of the type ``divev $=$
finite $+$ divev''.

Ultimately, the problem does not arise in the theories we are considering
here, for the following reasons. Both the classical action and counterterms
are local functionals, equal to integrals of local functions of dimension 4.
In the paper we also show that the first nonvanishing contributions to the
anomaly functional (\ref{anom}) are local, equal to integrals of local
functions of dimension 5. A fermion bilinear $\bar{\psi}_{1}\gamma ^{\rho
_{1}\cdots \rho _{k}}\psi _{2}$ has dimension 3, so power counting implies
that the classical action, as well as counterterms and local contributions
to anomalies, cannot contain products of two or more fermion bilinears.
Therefore, they are not affected by the ambiguities discussed above. Those
ambiguities can only occur in the convergent sector of the theory, where
they are harmless, since both analytic and formal evanescences must
eventually disappear.

Thanks to the properties just mentioned, it is meaningful to require that
the action $S_{r0}$, as well as its extensions constructed in the rest of
this paper, do not contain analytically evanescent terms. More precisely,
the coefficients of every Lagrangian terms should be equal to their
four-dimensional limits. This request is important to avoid unwanted
simplifications between $\varepsilon $ factors and $\varepsilon $ poles,
when divergent parts are extracted from bilinear expressions such as $%
(\Gamma ,\Gamma )$. It can be considered part of the definition of the
minimal subtraction scheme. For the same reason, we must be sure that the
antiparentheses do not generate extra factors of $\varepsilon $, or poles in 
$\varepsilon $, which is proved below.

Finite nonevanescent contributions will be called ``nev''. We need a
convention to define these quantities precisely, otherwise they can mix with
evanescent terms. For example, we need to state whether $\bar{C}\partial
^{2}C$, or $\bar{C}\bar{\partial}^{2}C$, or a combination such as $(1+\alpha
\varepsilon )\bar{C}\bar{\partial}^{2}C+\beta \bar{C}\hat{\partial}^{2}C$,
where $\alpha $ and $\beta $ are constants, is taken to be nonevanescent.
The convention we choose is that nonevanescent terms are maximally symmetric
with respect to the $D$-dimensional Lorentz group. For the arguments of this
paper we just need to focus on local functionals contributing to
counterterms and anomalies. In the case of counterterms the nonevanescent
terms are those appearing in the action $S_{r0}$, which are $SO(D)$%
-invariant when chiral fermions are switched off. In the case of anomalies
the nonevanescent terms are $SO(D)$-invariant unless they contain the tensor 
$\varepsilon ^{\mu \nu \rho \sigma }$ or chiral fermions.

\subsection{Evanescent extension of the classical action}

It is convenient to extend the action $S_{r0}$ by adding all formally
evanescent terms that have the features of divergent evanescences,
multiplied by independent parameters $\eta $. In this way it is possible to
subtract divergent evanescences by means of $\eta $ redefinitions. Denoting
the correction collecting such terms with $S_{\text{ev}}$, the extended
action reads 
\begin{equation}
S_{r}(\Phi ,K)=S_{r0}(\Phi ,K)+S_{\text{ev}}(\Phi ,K)=S_{c}+S_{\text{LR}}+S_{%
\text{ev}}+(S_{K},\Psi )+S_{K}=S_{\text{gf}}+S_{\text{LR}}+S_{\text{ev}}.
\label{exta}
\end{equation}%
Then the generating functionals (\ref{zg}), the functional $\Gamma $ and the
anomaly functional $\mathcal{A}$ of (\ref{anom}) are turned into those
defined by $S_{r}$.

Each term of $S_{\text{ev}}$ is the integral of a monomial of dimension $%
\leqslant 4$, globally invariant under $G$. It not necessarily gauge
invariant, since gauge invariance is violated away from four dimensions.
Moreover, $S_{\text{ev}}$ is $B$, $K_{B}$, $K_{\bar{C}}$, $\bar{\psi}_{R}$
and $\psi _{R}$ independent, linear in $K$ and depends on $\bar{C}^{a_{i}}$
and the sources $K^{\mu a_{i}}$ only through the combinations $K^{\mu
a_{i}}+\partial ^{\mu }\bar{C}^{a_{i}}$. It is also independent of $K_{C}$, $%
K_{\psi }$, $\bar{K}_{\psi }$, $\psi _{L}$ and $\bar{\psi}_{L}$, because no
formally evanescent terms can be built with these objects. By power counting
and ghost-number conservation the terms proportional to $K^{\mu
a_{i}}+\partial ^{\mu }\bar{C}^{a_{i}}$ are independent of matter fields. In
the end, $S_{\text{ev}}$ has the form 
\begin{equation}
S_{\text{ev}}(\Phi ,K)=S_{c\hspace{0.01in}\text{ev}}(A)-\int \sum_{i}R_{\mu 
\hspace{0.01in}\text{ev}}^{a_{i}}(A,C)(K^{\mu a_{i}}+\partial ^{\mu }\bar{C}%
^{a_{i}}).  \label{sev}
\end{equation}

We can further restrict $S_{\text{ev}}$. Indeed, $S_{r0}$ satisfies (\ref%
{liable}). Therefore, the divergent evanescences have the form (\ref{liable2}%
) with $L\geqslant 1$, and can be renormalized with an $S_{\text{ev}}$ of
the form (\ref{liable}). Precisely, we can define the parameters $\eta $ so
that $S_{\text{ev}}$ is linear in $\eta $ and its $g$ dependence has the
form 
\begin{equation}
S_{\text{ev}}(\Phi ,K,g,\eta )=\frac{1}{g^{2}}S_{\text{ev}}^{\prime }(g\Phi
,gK,\eta )\equiv \frac{1}{g^{2}}S_{c\hspace{0.01in}\text{ev}}^{\prime
}(gA,\eta )-\frac{1}{g^{2}}\int \sum_{i}R_{\mu \hspace{0.01in}\text{ev}%
}^{a_{i}\hspace{0.01in}\prime }(gA,gC,\eta )(gK^{\mu a_{i}}+g\partial ^{\mu }%
\bar{C}^{a_{i}}),  \label{eta}
\end{equation}%
so $S_{r}$ also satisfies (\ref{liable}).

Basically, the terms of $S_{\text{ev}}$ are similar to those appearing in $%
S_{r0}$, but contain some evanescent components of momenta and/or gauge
fields, and are broken into gauge noninvariant pieces. We have 
\begin{equation}
R_{\mu \hspace{0.01in}\text{ev}}^{a_{i}}=\eta _{1i}\partial _{\hat{\mu}%
}C^{a_{i}}+\eta _{2i}gf^{a_{i}b_{i}c_{i}}A_{\hat{\mu}}^{b_{i}}C^{c_{i}},
\label{rev}
\end{equation}%
while examples of contributions to $S_{c\hspace{0.01in}\text{ev}}$ are 
\begin{eqnarray}
S_{c\hspace{0.01in}\text{ev}} &=&\sum_{i}\int \left( \eta _{3i}(\partial
_{\mu }A_{\hat{\nu}}^{a_{i}})(\partial ^{\mu }A^{\hat{\nu}a_{i}})+\eta
_{4i}(\partial _{\hat{\mu}}A_{\nu }^{a_{i}})(\partial ^{\hat{\mu}}A^{\nu
a_{i}})+\eta _{5i}(\partial _{\hat{\mu}}A_{\hat{\nu}}^{a_{i}})(\partial ^{%
\hat{\mu}}A^{\hat{\nu}a_{i}})\right)  \notag \\
&&+\sum_{i}\int \left( \eta _{6i}(\partial _{\hat{\mu}}A^{\hat{\mu}%
a_{i}})(\partial _{\nu }A^{\nu a_{i}})+\eta _{7i}(\partial _{\hat{\mu}}A^{%
\hat{\mu}a_{i}})(\partial _{\hat{\nu}}A^{\hat{\nu}a_{i}})+\eta _{8i}A_{\hat{%
\mu}}^{a_{i}}A^{\hat{\mu}a_{i}}\right)  \label{scev} \\
&&+\sum_{i}\int \left( \eta _{9i}gf^{a_{i}b_{i}c_{i}}A_{\mu }^{a_{i}}A_{\hat{%
\nu}}^{b_{i}}\partial ^{\mu }A^{\hat{\nu}c_{i}}+\cdots \right) .  \notag
\end{eqnarray}%
The terms multiplied by $\eta _{3i},\cdots \eta _{8i}$ are quadratic and
modify the propagators of the gauge fields $A_{\mu }^{a_{i}}$ and the
Lagrange multipliers $B^{a_{i}}$. We do not need to report here the modified
propagators, which are rather involved. We have checked, with the help of a
computer program, that they satisfy the requirements we need. In particular,
if $k$ denotes their momentum, ($i$) they are regular when any evanescent
components $\hat{k}$ of $k$ are set to zero; ($ii$) when the propagators are
differentiated with respect to any components $\bar{k}$, $\hat{k}$, or to
parameters of positive dimensions (such as $\eta _{8i}$), their behaviors
for large $k^{2}$ improve by at least one power; ($iii$) they have a regular
infrared behavior, which corresponds to the decoupling of the evanescent
components $A_{\hat{\mu}}^{a_{i}}$. Finally, their denominators are $%
SO(1,3)\times SO(-\varepsilon )$ scalars, like the denominators of the
fermion propagators (\ref{fpar}).

The extended action (\ref{exta}) satisfies 
\begin{equation*}
(S_{r},S_{r})=(S_{r0},S_{r0})+\mathcal{O}(\eta )\mathcal{O}(\varepsilon )=%
\mathcal{O}(\varepsilon )+\mathcal{O}(\eta )\mathcal{O}(\varepsilon ),
\end{equation*}
where $(S_{r0},S_{r0})$ is given by (\ref{srsr}).

\subsection{Structure of correlation functions}

Now we analyze the evaluation of correlation functions. We use the same
notation for a function and its Fourier transform, since no confusion is
expected to arise.

In momentum space, the terms of the classical action can be written in the
form 
\begin{equation}
\int \left( \prod_{i=1}^{n+r}\frac{\mathrm{d}^{D}k_{i}}{(2\pi )^{D}}\right)
\Phi ^{\alpha _{1}}(k_{1})\cdots \Phi ^{\alpha _{n}}(k_{n})K_{\beta
_{1}}(k_{n+1})\cdots K_{\beta _{r}}(k_{n+r})\hspace{0.01in}T_{\mu _{1}\cdots
\mu _{p}\alpha _{1}\cdots \alpha _{n}}^{\beta _{1}\cdots \beta _{r}}G^{\mu
_{1}\cdots \mu _{p}}(k_{1},\cdots ,k_{n+r}),  \label{corro}
\end{equation}
where $k_{1},\cdots ,k_{n+r}$ are the external momenta.The constants $T_{\mu
_{1}\cdots \mu _{p}\alpha _{1}\cdots \alpha _{n}}^{\beta _{1}\cdots \beta
_{r}}$ collect all tensors $\eta _{\mu \nu }$, $\varepsilon _{\mu \nu \rho
\sigma }$, $\delta _{\hat{\mu}\hat{\nu}}$, $\gamma $ matrices, structure
constants $f^{abc}$ and matrices $T^{a}$. In particular, every projector
onto hat components of momenta, fields and sources is moved inside $T_{\mu
_{1}\cdots \mu _{p}\alpha _{1}\cdots \alpha _{n}}^{\beta _{1}\cdots \beta
_{r}}$. Momentum conservation ensures that 
\begin{equation}
G^{\mu _{1}\cdots \mu _{p}}(k_{1},\cdots ,k_{n+r})=(2\pi )^{D}\delta
^{(D)}(P)\hspace{0.01in}\tilde{G}^{\mu _{1}\cdots \mu _{p}}(k_{1},\cdots
,k_{n+r}),\qquad P=\sum_{i=1}^{n+r}k_{i},  \label{consa}
\end{equation}
where the tensors $\tilde{G}^{\mu _{1}\cdots \mu _{p}}$ are polynomials that
depend on $n+r-1$ external momenta.

Propagators can be decomposed as sums of terms of the form 
\begin{equation}
T_{\mu _{1}\cdots \mu _{p}\alpha _{1}\alpha _{2}}^{\prime }\frac{N_{\text{%
prop}}^{\mu _{1}\cdots \mu _{p}}(k)}{D_{\text{prop}}(k)},  \label{propastru}
\end{equation}%
where $T_{\mu _{1}\cdots \mu _{p}\alpha _{1}\alpha _{2}}^{\prime }$ is a
constant tensor, $N_{\text{prop}}^{\mu _{1}\cdots \mu _{p}}(k)$ is a
polynomial $SO(1,D-1)$ tensor, and $D_{\text{prop}}(k)$ is a polynomial $%
SO(1,3)\times SO(-\varepsilon )$ scalar. The reason why $D_{\text{prop}}(k)$
is not fully $SO(1,D-1)$ invariant is that the regularized propagators do
not have $SO(1,D-1)$-scalar denominators, due to the parameters $\varsigma
_{I}$ of formula (\ref{fpar}) and the parameters $\eta $ provided by the
extension $S_{r0}\rightarrow S_{r}$ discussed above.

The Feynman diagrams of $\Gamma $ and $\mathcal{A}$ have structures
inherited from the structures (\ref{corro}) and (\ref{propastru}) of the
vertices and propagators. They can be written as sums of contributions of
the form (\ref{corro}), with tensors $G^{\mu _{1}\cdots \mu _{p}}$ that
satisfy (\ref{consa}), but now $\tilde{G}^{\mu _{1}\cdots \mu _{p}}$ are
integrals over internal momenta $p$ of rational functions 
\begin{equation}
\frac{N^{\mu _{1}\cdots \mu _{p}}(p,k)}{D(p,k)},  \label{ratio}
\end{equation}%
where the polynomial $N^{\mu _{1}\cdots \mu _{p}}(p,k)$ appearing in the
numerator is an $SO(1,D-1)$ tensor, and the polynomial $D(p,k)$ appearing in
the denominator is an $SO(1,3)\times SO(-\varepsilon )$ scalar. At $%
\varsigma _{IJ}=\delta _{IJ}$, $\eta =0$ the integrals $\tilde{G}^{\mu
_{1}\cdots \mu _{p}}$ are full $SO(1,D-1)$ tensors. Note that $\tilde{G}%
^{\mu _{1}\cdots \mu _{p}}$ have a regular limit when the evanescent
components $\hat{k}$ of the external momenta $k$ tend to zero.

For example, we can write 
\begin{equation}
\int \frac{\mathrm{d}^{D}p}{(2\pi )^{D}}\frac{(\hat{p}^{2})^{2}}{(\bar{p}%
^{2}+\varsigma _{I}^{2}\hat{p}^{2}-m^{2})^{2}((p-k)^{2}-m^{2})}=\delta _{%
\hat{\mu}\hat{\nu}}\delta _{\hat{\rho}\hat{\sigma}}\tilde{G}^{\mu \nu \rho
\sigma }(k,m),  \label{exa}
\end{equation}
where 
\begin{equation*}
\tilde{G}^{\mu \nu \rho \sigma }(k,m)=\int \frac{\mathrm{d}^{D}p}{(2\pi )^{D}%
}\frac{p^{\mu }p^{\nu }p^{\rho }p^{\sigma }}{(\bar{p}^{2}+\varsigma _{I}^{2}%
\hat{p}^{2}-m^{2})^{2}((p-k)^{2}-m^{2})}.
\end{equation*}
Then we include $\delta _{\hat{\mu}\hat{\nu}}\delta _{\hat{\rho}\hat{\sigma}%
} $ inside the constants $T_{\mu _{1}\cdots \mu _{p}\alpha _{1}\cdots \alpha
_{n}}^{\beta _{1}\cdots \beta _{r}}$. The remaining completely symmetric
tensor $\tilde{G}^{\mu \nu \rho \sigma }(k,m)$ is an integral with the
properties listed above.

It may be useful to write (\ref{corro}) in the more compact form 
\begin{equation}
\int L_{\mu _{1}\cdots \mu _{p}}(\Phi ,K)\hspace{0.01in}G^{\mu _{1}\cdots
\mu _{p}}(k_{1},\cdots ,k_{n+r}),  \label{decompo}
\end{equation}%
and then organize the expressions $L_{\mu _{1}\cdots \mu _{p}}(\Phi ,K)$ by
using the basis of fermion bilinears $\bar{\psi}_{1}\gamma ^{\rho _{1}\cdots
\rho _{k}}\psi _{2}$, and explicitly evaluate traces of spinor indices and
contractions of Lorentz indices. At the end, all Lorentz indices appear in
gauge fields, fermion bilinears, the tensor $\varepsilon _{\mu \nu \rho
\sigma }$ (if present) and $G^{\mu _{1}\cdots \mu _{p}}$, and are contracted
among one another, possibly after projections onto bar or hat components.

It is also convenient to expand 
\begin{equation}
G^{\mu _{1}\cdots \mu _{p}}(k)=\sum_{i}\Pi _{i}^{\mu _{1}\cdots \mu
_{p}}(k)G_{i}(k)=(2\pi )^{D}\delta ^{(D)}(P)\hspace{0.01in}\sum_{i}\Pi
_{i}^{\mu _{1}\cdots \mu _{p}}(k)\tilde{G}_{i}(k),  \label{consi}
\end{equation}%
where $G_{i}(k)$ and $\tilde{G}_{i}(k)$ are $SO(1,3)\times SO(-\varepsilon )$
scalars, and $\Pi _{i}^{\mu _{1}\cdots \mu _{p}}(k)$ are polynomials
constructed with $\eta _{\mu \nu }$, $\varepsilon _{\mu \nu \rho \sigma }$, $%
\delta _{\hat{\mu}\hat{\nu}}$ and the $n+r-1$ independent momenta $k$. Then
we can write the contribution (\ref{decompo}) to $\Gamma $ or $\mathcal{A}$
as 
\begin{equation}
\int L_{i}G^{i},  \label{ultra}
\end{equation}%
where 
\begin{equation*}
L_{i}=L_{\mu _{1}\cdots \mu _{p}}(\Phi ,K)\Pi _{i}^{\mu _{1}\cdots \mu
_{p}}(k)
\end{equation*}%
are also $SO(1,3)\times SO(-\varepsilon )$ scalars. After these operations,
the Lorentz indices appear in gauge fields, fermion bilinears, momenta $k$
and the tensor $\varepsilon _{\mu \nu \rho \sigma }$. They are contracted
among themselves, possibly after projections onto bar or hat components. At
this point, traces and index contractions must be evaluated explicitly,
because they may produce factors $\varepsilon $, which are important for the
expansions and limits that we are going to define.

The \textit{analytic expansion} around $\varepsilon =0$ of (\ref{decompo})
or (\ref{ultra}) is defined by expanding the scalars $G^{i}(k)$ in powers of 
$\varepsilon $ without affecting the evanescent components of external
momenta. The \textit{analytic limit} is the order zero of the analytic
expansion, once the poles in $\varepsilon $ have been subtracted away. The 
\textit{formal limit} $\varepsilon \rightarrow 0$ is the limit where the
evanescent components of gauge fields, external momenta and fermion
bilinears are dropped. The \textit{limit} $\varepsilon \rightarrow 0$ is the
analytic limit followed by the formal limit.

For the reasons explained above, the analytic and formal limits may be
ambiguous in the convergent sector of the theory, but they are unambiguous
in the divergent sector. More importantly, the limit $\varepsilon
\rightarrow 0$ is always unambiguous. Since the tensors $G^{\mu _{1}\cdots
\mu _{p}}$ are regular when any evanescent components $\hat{k}$ of the
external momenta $k$ are set to zero, the formal limits of (\ref{decompo})
and (\ref{ultra}) are well-defined.

When we use the expressions ``$\mathcal{O}(\varepsilon )$'' or ``ev'' we
mean any quantity that vanishes in the limit $\varepsilon \rightarrow 0$.
Clearly, ev $=$ aev $+$ fev.

\subsection{Locality of counterterms}

Now we comment on the locality of counterterms. The forms of the regularized
propagators ensure that a sufficient number of derivatives with respect to
physical $\bar{k}$ and/or evanescent $\hat{k}$ components of external
momenta $k$ kills the overall divergences of Feynman diagrams. If we
subtract the divergent evanescences, together with the ordinary divergences,
up to some order $n$, then both ordinary divergences and divergent
evanescences of order $n+1$ are polynomial in $\bar{k}$ and $\hat{k}$. The $%
S_{r0}$-extension $S_{r}=S_{r0}+S_{\text{ev}}$ of formula (\ref{exta})
allows us to subtract all of them in a way that is efficient for the proof
of the Adler-Bardeen theorem.

To complete the analysis it is useful to describe what happens if for some
reason we do not subtract some divergent evanescences. We use the
abbreviations \textquotedblleft loc\textquotedblright\ and \textquotedblleft
nl\textquotedblright\ to denote local and nonlocal contributions,
respectively. At one loop we miss counterterms of the form 
\begin{equation}
\hbar \frac{\text{loc fev}}{\varepsilon }.  \label{h1}
\end{equation}%
Consequently, at two loops we also miss counterterms for subdivergences.
Using the vertex (\ref{h1}) inside one-loop diagrams we get contributions of
the form 
\begin{equation}
\hbar ^{2}\left( \frac{\text{loc nev}}{\varepsilon }+\frac{\text{loc fev}}{%
\varepsilon }+\text{nl}\right) +\hbar ^{2}\left( \frac{\text{loc fev}}{%
\varepsilon ^{2}}+\frac{\text{nl fev}}{\varepsilon }+\text{nl fev}\right) .
\label{h12}
\end{equation}%
The first three terms are generated when the formal evanescence enters the
diagram, is converted into a factor $\varepsilon $ and simplifies a pole in $%
\varepsilon $. Symbolically, we express this occurrence (which is the basic
mechanism that originates potential anomalies) as 
\begin{equation}
\text{fev}\cap \text{one-loop}\rightarrow \hbar \left( \text{loc nev}+\text{%
loc fev}+\mathcal{O}(\varepsilon )\text{ nl}\right) .  \label{fevs}
\end{equation}%
The last three terms of (\ref{h12}) describe what happens when the formal
evanescence remains outside the diagram.

The first term of (\ref{h12}) must be subtracted, so the missing
counterterms at two loops are 
\begin{equation}
\hbar ^{2}\frac{\text{loc fev}}{\varepsilon ^{2}},\qquad \hbar ^{2}\frac{%
\text{loc fev}}{\varepsilon },\qquad \hbar ^{2}\frac{\text{nl fev}}{%
\varepsilon }.  \label{h2}
\end{equation}%
Even if the last term of this list is nonlocal, we still have no problem,
since the residues of the poles in $\varepsilon $ are formally evanescent.
However, when we use the first and third terms of (\ref{h2}) inside one-loop
diagrams, the formal evanescence can simplify another pole, by the mechanism
(\ref{fevs}), and give 
\begin{equation*}
\hbar ^{3}\frac{\text{nl nev}}{\varepsilon }+\hbar ^{3}\frac{\text{nl fev}}{%
\varepsilon ^{2}}+\hbar ^{3}\frac{\text{nl fev}}{\varepsilon }+\hbar ^{3}%
\text{nl}
\end{equation*}%
plus local poles. We see that nonlocal, nonevanescent divergences appear at
three loops. These are only partially compensated by analogous contributions
originated by the subtraction of the first term of (\ref{h12}). Those due to
the first term of (\ref{h2}), in particular, do not seem to disappear.

On the other hand, it is safe to subtract the divergent evanescences order
by order, together with nonevanescent divergences. In this paper we adopt
this prescription.

\subsection{Properties of the antiparentheses}

Now we study how divergences and evanescences propagate through the
antiparentheses. Indeed, in the proofs of renormalizability to all orders
and the Adler-Bardeen theorem, it is necessary to extract divergent parts of
antiparentheses such as $\mathcal{A}=(\Gamma ,\Gamma )$ or $(\Gamma ,%
\mathcal{A})$. This operation is not as simple as it sounds, because we must
be sure that the antiparentheses themselves do not generate poles or factors
of $\varepsilon $, in order to be able to say that, for example, the
divergent part of $(S_{r},\Gamma ^{(1)})$ is equal to $(S_{r},\Gamma _{\text{%
div}}^{(1)})$, where $\Gamma ^{(1)}$ it the one-loop contribution to $\Gamma 
$ and $\Gamma _{\text{div}}^{(1)}$ is the divergent part of $\Gamma ^{(1)}$.
Specifically, we prove that

($i$) \textit{the antiparentheses }$(X_{\text{conv}},Y_{\text{conv}})$%
\textit{\ of convergent functionals }$X_{\text{conv}}$ and $Y_{\text{conv}}$%
\textit{\ are convergent;}

($ii$)\textit{\ the antiparentheses }$(X_{\text{conv}},Y_{\text{ev}})$%
\textit{\ of convergent functionals }$X_{\text{conv}}$\textit{\ and
evanescent functionals }$Y_{\text{ev}}$ \textit{are evanescent;}

($iii$) \textit{the antiparentheses }$(X,Y)$\textit{\ do not generate either
poles in }$\varepsilon $\textit{\ or factors of }$\varepsilon $\textit{\ if }%
$X$\textit{, }$Y$\textit{\ and }$(X,Y)$\textit{\ do not involve products of
two or more fermion bilinears.}

For the uses we have in mind it is convenient to rephrase property ($iii$)
more explicitly as

($iii^{\prime }$) \textit{the antiparentheses }$(X_{\text{A}},Y_{\text{B}})$ 
\textit{of functionals }$X_{\text{A}}$\textit{\ and }$Y_{\text{B}}$\textit{\
with the properties specified by their subscripts }A\textit{\ and }B\textit{%
, satisfy the identities} 
\begin{eqnarray}
(X_{\text{fev}},Y_{\text{nev/fev}}) &=&\text{fev},\qquad (X_{\text{divev}%
},Y_{\text{nev/fev/divev}})=\text{divev},\qquad (X_{\text{ev}},Y_{\text{fev}%
})=\text{ev},  \notag \\
\left. (X_{\text{nev}},Y_{\text{div}})\right| _{\text{div}} &=&(X_{\text{nev}%
},Y_{\text{div}}),\qquad \left. (X_{\text{nev}},Y_{\text{nev}})\right| _{%
\text{nev}}=(X_{\text{nev}},Y_{\text{nev}}),  \label{loco} \\
\left. (X_{\text{nev}},Y_{\text{nev\hspace{0.01in}div}})\right| _{\text{nev%
\hspace{0.01in}div}} &=&(X_{\text{nev}},Y_{\text{nev\hspace{0.01in}div}}), 
\notag
\end{eqnarray}
\textit{as long as }$X_{\text{A}}$, $Y_{\text{B}}$\textit{\ and }$(X_{\text{A%
}},Y_{\text{B}})$\textit{\ do not involve products of two or more fermion
bilinears.}

To prove these properties it is convenient to write the antiparentheses in
momentum space. We have 
\begin{equation}
\int \mathrm{d}^{D}x\frac{\delta _{r}X}{\delta \Phi ^{\alpha }(x)}\frac{%
\delta _{l}Y}{\delta K_{\alpha }(x)}=\int \frac{\mathrm{d}^{D}p}{(2\pi )^{D}}%
\frac{\delta _{r}X}{\delta \Phi ^{\alpha }(p)}\frac{\delta _{l}Y}{\delta
K_{\alpha }(-p)}  \label{inte}
\end{equation}%
and a similar relation obtained by exchanging $\Phi $ and $K$. Let us write
formulas (\ref{ultra}) for $X$, $Y$ and $(X,Y)$ as 
\begin{equation*}
X=\int L_{iX}G_{X}^{i},\qquad Y=\int L_{jY}G_{Y}^{j},\qquad (X,Y)=\int
L_{ij(X,Y)}G_{XY}^{ij}.
\end{equation*}%
Using (\ref{consi}) we find that the $p$ integral of formula (\ref{inte})
can be readily done and gives 
\begin{equation*}
G_{XY}^{ij}=(2\pi )^{D}\delta ^{(D)}(P)\hspace{0.01in}\tilde{G}_{X}^{i}%
\tilde{G}_{Y}^{j},
\end{equation*}%
where $P$ is the total momentum of $\tilde{G}_{X}^{i}$ plus the one of $%
\tilde{G}_{Y}^{j}$. We see that the scalar \textquotedblleft
cores\textquotedblright\ $G^{i}$ of correlation functions just multiply each
other in momentum space, which cannot generate new poles in $\varepsilon $
or factors of $\varepsilon $.

It remains to study the relation between $L_{ij(X,Y)}$ and $L_{iX}$, $L_{jY}$%
. The antiparentheses can produce index contractions by means the paired
functional derivatives $\delta /\delta A_{\mu }$-$\delta /\delta K^{\mu }$
and $\delta /\delta \psi $-$\delta /\delta K_{\psi }$. Clearly, no such
operations can generate poles in $\varepsilon $. This observation is
sufficient to prove statements ($i$) and ($ii$).

As far as statement ($iii$) is concerned, we must assume that the
functionals $X$, $Y$ and $(X,Y)$ do not involve products of two or more
fermion bilinears. Therefore, they are free of ambiguities of type (\ref%
{divam}). The contraction of Lorentz indices brought by $\delta /\delta
A_{\mu }$ and $\delta /\delta K^{\mu }$ gives a tensor $\eta ^{\mu \nu }$
with mixed indices (namely one index from $X$ and one index from $Y$). The
contraction of spinorial indices brought by $\delta /\delta \psi $ and $%
\delta /\delta K_{\psi }$ gives structures such as 
\begin{equation*}
\bar{\psi}_{1}\gamma ^{\rho _{1}\cdots \rho _{k}}\gamma ^{\sigma _{1}\cdots
\sigma _{l}}\psi _{2},
\end{equation*}%
where the $\rho $ indices come from $X$ and the $\sigma $ indices come from $%
Y$. Anticommuting the $\gamma $'s we can rearrange the indices so that $\rho
_{1}<\rho _{2}<\cdots <\rho _{k}$ and $\sigma _{1}<\sigma _{2}<\cdots
<\sigma _{l}$. Reordering the indices we may get minus signs from further
anticommutations or from squares of $\gamma $ matrices with identical
indices. In the end, we get a formula like 
\begin{equation*}
\bar{\psi}_{1}\gamma ^{\rho _{1}\cdots \rho _{k}}\gamma ^{\sigma _{1}\cdots
\sigma _{l}}\psi _{2}=\sum \pm \bar{\psi}_{1}\gamma ^{\rho _{1}\cdots \breve{%
\rho}_{m}\cdots \rho _{k}\sigma _{1}\cdots \breve{\sigma}_{n}\cdots \sigma
_{l}}\psi _{2}\prod \eta ^{\rho _{m}\sigma _{n}},
\end{equation*}%
where the breves denote missing indices that go into the tensors $\eta ^{\mu
\nu }$. Again, we get only tensors $\eta ^{\mu \nu }$ with mixed indices. We
recall that all Lorentz indices, possibly after projection onto bar or hat
components, are contracted with gauge fields, fermion bilinears, momenta and
possibly $\varepsilon _{\mu \nu \rho \sigma }$, and that, by assumption, no
products of two or more fermion bilinears are involved. Then it is obvious
that the contractions originated by the antiparentheses cannot produce $%
\varepsilon $ factors. Using these properties it is easy to check that
identities (\ref{loco}) hold, so statement ($iii$) is also proved.

Statement ($iii$) also says that the antiparentheses cannot convert formal $%
\varepsilon $ evanescences into analytic ones. It applies, for example, to
local functionals $X$ and $Y$ that are equal to the integrals of functions
of dimensions $n_{X},n_{Y}$ $\leqslant 5$, such that $n_{X}+n_{Y}$ $%
\leqslant 8$, because then $X$, $Y$ and $(X,Y)$ cannot contain products of
two or more fermion bilinears. In the paper we will apply statement ($iii$)
to the divergent contributions to $\Gamma $ and the first nonvanishing
contributions to the anomaly functional $\mathcal{A}$ of (\ref{an}).

\section{DHD regularization}

\setcounter{equation}{0}

The dimensional regularization alone does not provide the subtraction scheme
where the cancellation of gauge anomalies is manifest to all orders. To find
the right scheme, we modify the regularization technique by adding
higher-derivative terms that preserve gauge invariance in $D=4$. We take the
non-gauge-fixed regularized classical action 
\begin{equation}
S_{c\Lambda }=S_{c}+S_{\text{LR}}-\frac{1}{4}\int F_{\mu \nu }^{a}\left( 
\frac{D^{2}}{\Lambda ^{2}}\right) ^{\!\!8}F^{a\hspace{0.01in}\mu \nu
}+\sum_{I}\int \bar{\psi}_{L}^{I}\imath \slashed{D}\left( \frac{D^{2}}{%
\Lambda ^{2}}\right) ^{\!\!3}\psi _{L}^{I}+S_{\Lambda \text{LR}},
\label{lla}
\end{equation}%
where 
\begin{equation}
S_{\Lambda \text{LR}}=\sum_{I}\int \bar{\psi}^{I}\imath \slashed{\partial}%
\left( \frac{\partial ^{2}}{\Lambda ^{2}}\right) ^{\!\!3}\psi
^{I}-\sum_{I}\int \bar{\psi}_{L}^{I}\imath \slashed{\partial}\left( \frac{%
\partial ^{2}}{\Lambda ^{2}}\right) ^{\!\!3}\psi _{L}^{I}.  \label{lla2}
\end{equation}%
The higher-derivative structures of (\ref{lla}) and (\ref{lla2}) are chosen
to simplify the arguments of our derivations.

We gauge fix $S_{c\Lambda }$ using modified gauge-fixing functions of the
form 
\begin{equation}
\mathcal{G}_{\Lambda }^{a_{i}}=Q\left( \Box \right) \partial ^{\mu }A_{\mu
}^{a_{i}},\qquad Q\left( \Box \right) =1+\frac{\lambda ^{\prime }}{\Lambda
^{16}}\Box ^{8},  \label{ga}
\end{equation}%
and a modified gauge fermion 
\begin{equation*}
\Psi _{\Lambda }(\Phi )=\sum_{i}\int \bar{C}^{a_{i}}\left( \mathcal{G}%
_{\Lambda }^{a_{i}}+\frac{1}{2}P_{i}\left( \Box \right) B^{a_{i}}\right)
,\qquad P_{i}\left( \Box \right) =\xi _{i}+\frac{\xi ^{\prime }}{\Lambda
^{16}}\Box ^{8},
\end{equation*}%
where $\lambda ^{\prime }$ and $\xi ^{\prime }$ are other (dimensionless)
gauge-fixing parameters.

Finally, we add 
\begin{equation*}
S_{\Lambda \hspace{0.01in}\text{ev}}=S_{c\hspace{0.01in}\text{ev}}(A)-\int
\sum_{i}R_{\mu \hspace{0.01in}\text{ev}}^{a_{i}}(A,C)\left( K^{\mu
a_{i}}+Q\left( \Box \right) \partial ^{\mu }\bar{C}^{a_{i}}\right) ,
\end{equation*}%
which differs from $S_{\text{ev}}$ only because the combinations $K^{\mu
a_{i}}+\partial ^{\mu }\bar{C}^{a_{i}}$ are replaced by $K^{\mu
a_{i}}+Q\left( \Box \right) \partial ^{\mu }\bar{C}^{a_{i}}$.

The regularized gauge-fixed action reads 
\begin{equation}
S_{\Lambda }(\Phi ,K)=S_{c\Lambda }+S_{\Lambda \hspace{0.01in}\text{ev}%
}+(S_{K},\Psi _{\Lambda })+S_{K},  \label{sl}
\end{equation}%
where $S_{K}$ is the same as before, and satisfies 
\begin{equation}
(S_{\Lambda },S_{\Lambda })=2g\int C^{a}\left( (\partial _{\hat{\mu}}\hspace{%
0.01in}h_{IJ}(\partial ^{2})\bar{\psi}_{R}^{I})\imath \gamma ^{\hat{\mu}%
}T^{a}\psi _{L}^{J}+\bar{\psi}_{L}^{I}T^{a}\imath \hat{\slashed{\partial}}%
\hspace{0.01in}h_{JI}^{\ast }(\partial ^{2})\psi _{R}^{J}\right) +\mathcal{O}%
(\eta )\mathcal{O}(\varepsilon ),  \label{srsr2}
\end{equation}%
where $h_{IJ}(\partial ^{2})=(\varsigma _{IJ}\Lambda ^{6}+\delta
_{IJ}(\partial ^{2})^{3})/\Lambda ^{6}$. The reason why it is useful to
separate the terms proportional to the parameters $\eta $ will become clear
later.

It is straightforward to derive the propagators and check that the ones of
gauge fields, $\langle A_{\mu }(k)\hspace{0.01in}A_{\nu }(-k)\rangle _{0}$,
and the ones of ghosts, $\langle C(k)\hspace{0.01in}\bar{C}(-k)\rangle _{0}$%
, fall off as $1/(k^{2})^{9}$ for large momenta $k$, while the propagators $%
\langle A(k)\hspace{0.01in}B(-k)\rangle _{0}$ fall off as $k/(k^{2})^{9}$,
and $\langle B(k)\hspace{0.01in}B(-k)\rangle _{0}$ as $1/(k^{2})^{8}$. For
example, in the \textquotedblleft Feynman gauge\textquotedblright\ $\xi
_{i}=\lambda ^{\prime }=\xi ^{\prime }=1$ at $\eta =0$ we have 
\begin{equation}
\langle A_{\mu }(k)\hspace{0.01in}A_{\nu }(-k)\rangle _{0}=-\frac{\imath
\eta _{\mu \nu }}{k^{2}Q(-k^{2})},\qquad \langle C(k)\hspace{0.01in}\bar{C}%
(-k)\rangle _{0}=\frac{\imath }{k^{2}Q(-k^{2})}.  \label{propa}
\end{equation}%
The fermion propagators, on the other hand, fall off as $p/(p^{2})^{4}$.

For a while we need to work at finite $\Lambda $, where the action $%
S_{\Lambda }$ is super-renormalizable. To make its super-renormalizability
manifest, it is convenient to parametrize it so that the $\Lambda $
denominators cancel out. Let us first ignore the terms $S_{\Lambda \hspace{%
0.01in}\text{ev}}$. We define tilde fields and tilde parameters as 
\begin{equation}
\tilde{A}_{\mu }^{a}=\frac{A_{\mu }^{a}}{\Lambda ^{8}},\qquad \tilde{\psi}%
^{I}=\frac{\psi ^{I}}{\Lambda ^{3}},\qquad \tilde{g}=\Lambda ^{8}g,\qquad 
\tilde{\zeta}_{i}=\Lambda ^{16}\zeta _{i},  \label{resca}
\end{equation}%
and $\tilde{r}_{i}=r_{i}$. The covariant derivatives remain $\Lambda $
in\-depen\-dent. To cancel the $\Lambda $ denominators of the gauge-fixing
sector we define $\widetilde{\bar{C}}^{a}=\bar{C}^{a}/\Lambda ^{8}$, $\tilde{%
B}^{a}=B^{a}/\Lambda ^{8}$ and $\tilde{C}^{a}=C^{a}/\Lambda ^{8}$. Finally,
we define the tilde sources 
\begin{equation*}
(\tilde{K}^{\mu a},\tilde{K}_{C}^{a},\tilde{K}_{\bar{C}}^{a},\tilde{K}%
_{B}^{a},\tilde{K}_{\psi }^{I},\widetilde{\bar{K}}_{\psi }^{I})=(\Lambda
^{8}K^{\mu a},\Lambda ^{8}K_{C}^{a},\Lambda ^{8}K_{\bar{C}}^{a},\Lambda
^{8}K_{B}^{a},\Lambda ^{3}K_{\psi }^{I},\Lambda ^{3}\bar{K}_{\psi }^{I}),
\end{equation*}%
so the tilde map is a canonical transformation combined with a redefinition
of parameters.

As far as $S_{\Lambda \text{ev}}$ is concerned, using (\ref{eta}) and the
linearity in $\eta $ we can write it as 
\begin{equation}
\tilde{S}_{\Lambda \hspace{0.01in}\text{ev}}=\frac{1}{\tilde{g}^{2}}S_{c%
\hspace{0.01in}\text{ev}}^{\prime }(\tilde{g}\tilde{A},\Lambda ^{16}\eta )-%
\frac{1}{\tilde{g}^{2}}\sum_{i}\int R_{\mu \hspace{0.01in}\text{ev}}^{a_{i}%
\hspace{0.01in}\prime }(\tilde{g}\tilde{A},\tilde{g}\tilde{C},\eta )\left( 
\tilde{g}\tilde{K}^{\mu a_{i}}+\tilde{g}\tilde{Q}\left( \Box \right)
\partial ^{\mu }\widetilde{\bar{C}}^{a_{i}}\right) ,  \label{ss}
\end{equation}%
where $\tilde{Q}\left( \Box \right) =\Lambda ^{16}+\lambda ^{\prime }\Box
^{8}$.

In the tilde parametrization the full action reads 
\begin{eqnarray}
\tilde{S}_{\Lambda }(\tilde{\Phi},\tilde{K}) &\equiv &S_{\Lambda }(\Phi (%
\tilde{\Phi}),K(\tilde{K}))=-\frac{1}{4}\sum_{i}\int \tilde{F}_{\mu \nu
}^{a_{i}}\left( \tilde{\zeta}_{i}+(\tilde{D}^{2})^{8}\right) \tilde{F}^{a_{i}%
\hspace{0.01in}\mu \nu }  \notag \\
&&+\int \widetilde{\bar{\psi}}_{L}^{I}\imath \widetilde{\slashed{D}}\left(
\Lambda ^{6}+(\tilde{D}^{2})^{3}\right) \tilde{\psi}_{L}^{I}+\int \widetilde{%
\bar{\psi}}_{R}^{I}\imath \slashed{\partial}\left( \Lambda ^{6}+(\partial
^{2})^{3}\right) \widetilde{\psi }_{R}^{I}  \notag \\
&&+\int \widetilde{\bar{\psi}}_{R}^{I}\imath \slashed{\partial}\left(
\varsigma _{IJ}\Lambda ^{6}+\delta _{IJ}(\partial ^{2})^{3}\right) 
\widetilde{\psi }_{L}^{J}+\int \widetilde{\bar{\psi}}_{L}^{I}\imath %
\slashed{\partial}\left( \varsigma _{JI}^{\ast }\Lambda ^{6}+\delta
_{IJ}(\partial ^{2})^{3}\right) \widetilde{\psi }_{R}^{J}  \notag \\
&&+\sum_{i}\int \tilde{B}^{a_{i}}\tilde{Q}\left( \Box \right) \partial ^{\mu
}\tilde{A}_{\mu }^{a_{i}}+\frac{1}{2}\sum_{i}\int \tilde{B}^{a_{i}}\tilde{P}%
_{i}\left( \Box \right) \tilde{B}^{a_{i}}-\sum_{i}\int \widetilde{\bar{C}}%
^{a_{i}}\tilde{Q}\left( \Box \right) \partial ^{\mu }\tilde{D}_{\mu }\tilde{C%
}^{a_{i}}  \notag \\
&&-\int R^{\alpha }(\tilde{\Phi},\tilde{g})\tilde{K}_{\alpha }+\tilde{S}%
_{\Lambda \hspace{0.01in}\text{ev}},  \label{stl}
\end{eqnarray}%
where $\tilde{P}_{i}\left( \Box \right) =\tilde{\xi}_{i}+\xi ^{\prime }\Box
^{8}$, $\tilde{\xi}_{i}=\xi _{i}\Lambda ^{16}$.

The DHD-regularized generating functional $Z_{\Lambda }$ reads 
\begin{equation*}
Z_{\Lambda }(J,K)=\int [\mathrm{d}\Phi ]\exp \left( \imath S_{\Lambda }(\Phi
,K)+\imath \int \Phi ^{\alpha }J_{\alpha }\right) =\exp \imath W_{\Lambda
}(J,K),
\end{equation*}%
and the generating functional $\Gamma _{\Lambda }(\Phi ,K)=W_{\Lambda
}(J,K)-\int \Phi ^{\alpha }J_{\alpha }$ of one-particle irreducible diagrams
is the Legendre transform of $W_{\Lambda }(J,K)$ with respect to $J$. Since
no one-particle irreducible diagrams with external legs $\psi _{R}$, $\bar{%
\psi}_{R}$ can be constructed, the action $S_{\Lambda }$ and the $\Gamma $
functional $\Gamma _{\Lambda }$ depend on $\psi _{R}$, $\bar{\psi}_{R}$ in
exactly the same way. The DHD-regularized anomaly functional is 
\begin{equation}
\mathcal{A}_{\Lambda }=(\Gamma _{\Lambda },\Gamma _{\Lambda })=\langle
(S_{\Lambda },S_{\Lambda })\rangle _{S_{\Lambda }}.  \label{an}
\end{equation}%
When we switch to the tilde parametrization we write $\tilde{Z}_{\Lambda }$, 
$\tilde{W}_{\Lambda }$, $\tilde{\Gamma}_{\Lambda }$ and $\mathcal{\tilde{A}}%
_{\Lambda }$. See appendix B for the proof of the last equality of (\ref{an}%
).

The tilde action $\tilde{S}_{\Lambda }$ is polynomial in $\Lambda $, has
properly normalized propagators and contains only parameters of nonnegative
dimensions in units of mass. However, the tilde fields have negative
dimensions, which in principle may jeopardize the (super)renormalizability
we want to prove. Precisely, we have 
\begin{equation*}
\lbrack \tilde{A}]=[\widetilde{\bar{C}}]=[\widetilde{C}]=-7,\qquad \lbrack 
\tilde{B}]=-6,\qquad \lbrack \tilde{\psi}]=-\frac{3}{2},\qquad \lbrack 
\tilde{g}]=8,
\end{equation*}%
while $[\tilde{K}^{a\hspace{0.01in}\mu }]=[\tilde{K}_{C}^{a}]=[\tilde{K}_{%
\bar{C}}^{a}]=10$, $[\tilde{K}_{B}]=9$ and $[\tilde{K}_{\psi }]=9/2$. The
problem is solved as follows. Since $S_{\Lambda }$ has the form (\ref{liable}%
), the $\tilde{g}$ structure of $\tilde{S}_{\Lambda }$ is the tilde version
of (\ref{liable}). The tilde version of formula (\ref{liable2}) ensures that
the counterterms have the $\tilde{g}$ structure 
\begin{equation}
\sum_{L\geqslant 1}\tilde{g}^{2(L-1)}F_{L}(\tilde{g}\tilde{\Phi},\tilde{g}%
\tilde{K}),  \label{counter}
\end{equation}%
where the $L$-loop local functionals $F_{L}$ depend polynomially on the
other dimensionful parameters of the theory. Then we see that the theory is
indeed superrenormalizable, because the dimensions of all products $\tilde{g}%
\tilde{\Phi}$ and $\tilde{g}\tilde{K}$ are strictly positive.

\subsection{The DHD\ limit}

The basic idea behind the DHD regularization is to \textquotedblleft first
send $\varepsilon $ to zero, then $\Lambda $ to infinity\textquotedblright .
However, we must formulate the rules of such limits more precisely, since
certain caveats demand attention. We distinguish the \textit{%
higher-derivative theory} from the \textit{final theory}. The
higher-derivative theory is the one defined by the classical action $%
S_{\Lambda }$ (or $\tilde{S}_{\Lambda }$, if we use the tilde
parametrization), where the scale $\Lambda $ is kept fixed and treated like
any other parameter, instead of a cutoff. It is super-renormalizable and
regularized by the dimensional technique. Its divergences, which are poles
in $\varepsilon $, are subtracted in the next section using the minimal
subtraction scheme. The final theory is obtained by taking the limit $%
\Lambda \rightarrow \infty $ on the renormalized higher-derivative theory,
after subtracting the $\Lambda $ divergences that emerge in that limit.

Having already expanded in $\varepsilon $, we may wonder what types of
divergences appear in the final theory. We have products $\Lambda ^{k}\ln
^{k^{\prime }}\Lambda $ of powers and logarithms of $\Lambda $, but we also
have terms that are evanescent in $\varepsilon $ and divergent in $\Lambda $%
. To understand what to do with these, we distinguish two types of them,
according to whether the $\varepsilon $ evanescence is analytic or formal.

($i$)\ First, consider analytic evanescences in $\varepsilon $ multiplied by
products $\Lambda ^{k}\ln ^{k^{\prime }}\Lambda $, such as $\varepsilon
\Lambda ^{2}\ln \Lambda $. Since we first send $\varepsilon $ to zero, these
quantities are not true divergences and must be neglected. In any case, they
cannot be subtracted away, because the theorem of locality of counterterms
does not apply to them. Consider for example the integral 
\begin{equation*}
\int \frac{\mathrm{d}^{D}p}{(2\pi )^{D}}\frac{\Lambda ^{4}}{%
(p^{2}+m^{2})(\Lambda ^{4}+(p^{2})^{2})}=\frac{\Lambda ^{4-\varepsilon }m^{2}%
\left[ \cos \left( \frac{\pi \varepsilon }{4}\right) +\frac{\Lambda ^{2}}{%
m^{2}}\sin \left( \frac{\pi \varepsilon }{4}\right) -\frac{\Lambda
^{\varepsilon }}{m^{\varepsilon }}\right] }{2^{D}\pi ^{(D-2)/2}\Gamma (\frac{%
D}{2})(\Lambda ^{4}+m^{4})\sin \left( \frac{\pi \varepsilon }{2}\right) },
\end{equation*}
where for the purposes of our present discussion the mass $m$ can also play
the role of an external momentum. Expanding the right-hand side in powers of 
$\varepsilon $ we find that the $\mathcal{O}(\varepsilon ^{0})$ terms, which
are equal to 
\begin{equation*}
\frac{1}{32\pi ^{2}}\left( \pi \Lambda ^{2}-2m^{2}\ln \frac{\Lambda ^{2}}{%
m^{2}}\right) +\mathcal{O}(\frac{m}{\Lambda }),
\end{equation*}
have a $\Lambda $-divergent part that is polynomial in $m$, as expected,
while the $\mathcal{O}(\varepsilon ^{1})$ terms have a $\Lambda $-divergent
part that contains expressions such as 
\begin{equation*}
\Lambda ^{2}\ln \frac{\Lambda ^{2}}{m^{2}},\qquad m^{2}\ln ^{2}\frac{\Lambda
^{2}}{m^{2}},
\end{equation*}
which are not polynomial in $m$.

($ii$) Next, consider formal evanescences times $\Lambda ^{k}\ln ^{k^{\prime
}}\Lambda $, such as $(\ln \Lambda )\hspace{0.01in}\partial _{\mu }A_{\hat{%
\nu}}\partial ^{\mu }A^{\hat{\nu}}$. These can (actually, \textit{must}, for
the reasons explained in subsection 2.5) be subtracted away (as long as
their coefficients are calculated at $\varepsilon =0$), because the form of
regularized propagators ensures that counterterms are polynomial in both
physical and evanescent components of external momenta and fields.

($iii$)\ Formally evanescent expressions multiplied by products $\Lambda
^{k}\ln ^{k^{\prime }}\Lambda $ and factors of $\varepsilon $ are just like
case ($i$)\ and should not be subtracted away.

($iv$) For completeness, we point out a forth type of $\varepsilon $%
-evanescent $\Lambda $ divergences, that is to say nonlocal contributions of
type ($ii$), which can appear as artifacts of inconvenient manipulations.
Precisely, because of the ambiguities encoded in formula (\ref{divam}) some
quantities of type ($i$) can be converted into nonlocal divergences of type (%
$ii$). These conversions should just be avoided. To this purpose, it is
sufficient to note that the structure (\ref{corro}) of diagrams and the
expansion of the integrals $G^{\mu _{1}\cdots \mu _{p}}$ only generate $%
\varepsilon $-evanescent $\Lambda $ divergences of types ($i$), ($ii$) and ($%
iii$). In the event that \textquotedblleft aev $\rightarrow $ fev
conversions\textquotedblright\ of type (\ref{divam}) are accidentally
applied, nonlocal divergences of type ($ii$) can just be ignored, because
they cannot mix with the local terms belonging to the power-counting
renormalizable sector and they are resummable into contributions of type ($i$%
).

To summarize, the $\Lambda $ divergences are equal to $\Lambda ^{k}\ln
^{k^{\prime }}\Lambda $ times local monomials of the fields, the sources and
their derivatives. From the point of view of the dimensional regularization,
those monomials may be nonevanescent or formally evanescent, and their
coefficients must be evaluated in the analytic limit $\varepsilon
\rightarrow 0$.

We can thus define the procedure with which we renormalize the final theory
and define the physical quantities. We call it \textit{the} \textit{DHD limit%
}. We still organize the contributions to $\Gamma $ and $\mathcal{A}$ in the
form (\ref{corro}). Referring to (\ref{decompo}) and (\ref{ultra}), the DHD\
limit is made of the analytic limit $\varepsilon \rightarrow 0$, followed by
the limit $\Lambda \rightarrow \infty $, followed by the formal limit $%
\varepsilon \rightarrow 0$. We also have the \textit{DHD expansion}, that is
to say the analytic expansion around $\varepsilon =0$ followed by the
expansion around $\Lambda =\infty $.

The three steps that define the DHD limit are unambiguous in the divergent
sector, which does not contain products of more than one fermion bilinears.
Instead, the first and third steps are ambiguous in the convergent sector.
What is important is that the DHD limit is also unambiguous in the
convergent sector.

It is useful to recapitulate the DHD\ limit in symbolic form. We first
expand around $\varepsilon =0$ at $\Lambda $ fixed, and find poles, finite
terms and evanescent terms: 
\begin{equation*}
\frac{1}{\varepsilon },\qquad \frac{\hat{\delta}}{\varepsilon },\qquad
\varepsilon ^{0},\qquad \hat{\delta}\varepsilon ^{0},\qquad \varepsilon
,\qquad \hat{\delta}\varepsilon .
\end{equation*}%
The symbols appearing in this list have the following meanings: $%
1/\varepsilon $ denotes any kinds of divergences in $\varepsilon $, $\hat{%
\delta}$ is any formally evanescent quantity, $\varepsilon ^{0}$ is any
quantity that is convergent and nonevanescent in the analytic limit $%
\varepsilon \rightarrow 0$, and $\varepsilon $ denotes any analytic
evanescence. After the expansion, we subtract the poles and remain with 
\begin{equation}
\varepsilon ^{0},\qquad \hat{\delta}\varepsilon ^{0},\qquad \varepsilon
,\qquad \hat{\delta}\varepsilon .  \label{survi}
\end{equation}%
The terms proportional to $\varepsilon $ vanish in the DHD limit. The terms $%
\hat{\delta}\varepsilon ^{0}$ also vanish in that limit, but for some time
we treat them together with the $\varepsilon ^{0}$ terms. Next, we study the 
$\Lambda $ dependence. Expanding the coefficients of every surviving terms (%
\ref{survi}) around $\Lambda =\infty $, we find 
\begin{eqnarray}
&&\varepsilon ^{0}\Lambda ,\qquad \hat{\delta}\varepsilon ^{0}\Lambda
,\qquad \varepsilon ^{0}\Lambda ^{0},\qquad \hat{\delta}\varepsilon
^{0}\Lambda ^{0},\qquad \frac{\varepsilon ^{0}}{\Lambda },\qquad \frac{\hat{%
\delta}\varepsilon ^{0}}{\Lambda },  \notag \\
&&\varepsilon \Lambda ,\qquad \hat{\delta}\varepsilon \Lambda ,\qquad
\varepsilon \Lambda ^{0},\qquad \hat{\delta}\varepsilon \Lambda ^{0},\qquad 
\frac{\varepsilon }{\Lambda },\qquad \frac{\hat{\delta}\varepsilon }{\Lambda 
},  \label{assu}
\end{eqnarray}%
where $\Lambda $ denotes any kind of $\Lambda $-divergent expression (such
as $\Lambda ^{k}\ln ^{k^{\prime }}\Lambda $, with $k,k^{\prime }\geqslant 0$
and $k+k^{\prime }>0$), while $\Lambda ^{0}$ is any $\Lambda $-convergent,
non-$\Lambda $-evanescent expression, and $1/\Lambda $ is any $\Lambda $%
-evanescent expression. Then we subtract the $\Lambda $ divergences of the
DHD limit, namely the terms of types $\varepsilon ^{0}\Lambda $ and $\hat{%
\delta}\varepsilon ^{0}\Lambda $. After that we remain with 
\begin{equation}
\varepsilon ^{0}\Lambda ^{0},\qquad \hat{\delta}\varepsilon ^{0}\Lambda
^{0},\qquad \frac{\varepsilon ^{0}}{\Lambda },\qquad \frac{\hat{\delta}%
\varepsilon ^{0}}{\Lambda },\qquad \varepsilon \Lambda ,\qquad \hat{\delta}%
\varepsilon \Lambda ,\qquad \varepsilon \Lambda ^{0},\qquad \hat{\delta}%
\varepsilon \Lambda ^{0},\qquad \frac{\varepsilon }{\Lambda },\qquad \frac{%
\hat{\delta}\varepsilon }{\Lambda }.  \label{survo}
\end{equation}%
At this point we are ready to take the DHD limit, which drops all
contributions of this list but the $\varepsilon ^{0}\Lambda ^{0}$ terms.

\section{Renormalization of the higher-derivative theory}

\setcounter{equation}{0}

\label{subase}

In this section and the next two we study the higher-derivative regularized
theory $\tilde{S}_{\Lambda }$, keeping $\Lambda $ fixed and (mostly) using
the tilde parametrization. We first work out the renormalization of the
theory, then study its one-loop anomalies and finally prove the anomaly
cancellation to all orders.

The counterterms (\ref{counter}) are local and largely constrained. We know
that $i$) they are independent of $\tilde{B}$, $\tilde{K}_{\bar{C}}$, $%
\tilde{K}_{B}$, $\tilde{\psi}_{R}$ and $\widetilde{\bar{\psi}}_{R}$ and $ii$%
) do not depend on antighosts $\widetilde{\bar{C}}^{a_{i}}$ and sources $%
\tilde{K}^{\mu a_{i}}$ separately, but only through the combinations $\tilde{%
K}^{\mu a_{i}}+\tilde{Q}(\Box )\partial ^{\mu }\widetilde{\bar{C}}^{a_{i}}$.
Indeed, we have arranged $S_{\Lambda \hspace{0.01in}\text{ev}}$ to preserve
these properties. Actually, we have chosen the higher-derivative structure
of $S_{\Lambda }$ to simplify the counterterms even more: $iii$)\ they
cannot depend on the sources $\tilde{K}$ and matter fields $\tilde{\psi}$,
because each product $\tilde{g}\tilde{K}$, $\tilde{g}\tilde{\psi}$ has
dimension greater than 4; $iv$)\ they cannot contain antighosts, because of
points ($ii$) and ($iii$); $v$) they cannot contain ghosts, because all
objects with negative ghost numbers are excluded by points ($iii$) and ($iv$%
); $vi$) they can only be one-loop, because each loop carries an extra
factor $\tilde{g}^{2}$, which has dimension 16. In the end, there can only
be one-loop divergences of the form 
\begin{equation}
\partial ^{2}(\tilde{g}\tilde{A})^{2},\qquad \partial (\tilde{g}\tilde{A}%
)^{3},\qquad (\tilde{g}\tilde{A})^{4}  \label{con2}
\end{equation}%
(where derivatives can act on any objects to their right), and those
obtained from these expressions by suppressing some $\tilde{g}\tilde{A}$'s
or derivatives.

The anomaly functional (\ref{an}), if nonvanishing and nontrivial (in a
sense specified below), is the anomaly of the higher-derivative theory. In
the tilde parametrization we have 
\begin{equation}
\mathcal{\tilde{A}}_{\Lambda }=(\tilde{\Gamma}_{\Lambda },\tilde{\Gamma}%
_{\Lambda })=\langle (\tilde{S}_{\Lambda },\tilde{S}_{\Lambda })\rangle _{%
\tilde{S}_{\Lambda }}.  \label{ana}
\end{equation}%
The one-loop contribution $\mathcal{\tilde{A}}_{\Lambda }^{(1)}$ is 
\begin{equation}
\mathcal{\tilde{A}}_{\Lambda }^{(1)}=2(\tilde{S}_{\Lambda },\tilde{\Gamma}%
_{\Lambda }^{(1)})=\left. \langle (\tilde{S}_{\Lambda },\tilde{S}_{\Lambda
})\rangle _{\tilde{S}_{\Lambda }}\right\vert _{\text{one-loop}},
\label{ana2}
\end{equation}%
where $\tilde{\Gamma}_{\Lambda }^{(1)}$ is the one-loop contribution to $%
\tilde{\Gamma}_{\Lambda }$. Using (\ref{loco}) and (\ref{sl}) we see that $(%
\tilde{S}_{\Lambda },\tilde{S}_{\Lambda })=$ fev. The right-hand side of (%
\ref{ana2}) collects one-loop Feynman diagrams containing insertions of
formally evanescent vertices. The formal evanescences can: ($a$) remain
attached to external legs and momenta, or ($b$) be turned into one or more
factors $\varepsilon $. In case ($a$) they give local divergent evanescences
plus nonlocal evanescences. In case ($b$) the factors $\varepsilon $ can
simplify a local divergent part and give local nonevanescent contributions,
in addition to (generically nonlocal) evanescences. Therefore, we can write 
\begin{equation}
\mathcal{\tilde{A}}_{\Lambda }^{(1)}=\mathcal{\tilde{A}}_{\Lambda \hspace{%
0.01in}\text{nev}}^{(1)}+\mathcal{\tilde{A}}_{\Lambda \hspace{0.01in}\text{%
divev}}^{(1)}+\mathcal{\tilde{A}}_{\Lambda \hspace{0.01in}\text{ev}}^{(1)},
\label{adivev}
\end{equation}%
where $\mathcal{\tilde{A}}_{\Lambda \hspace{0.01in}\text{nev}}^{(1)}$ is
local, convergent and nonevanescent, $\mathcal{\tilde{A}}_{\Lambda \hspace{%
0.01in}\text{divev}}^{(1)}$ is local and divergent-evanescent and $\mathcal{%
\tilde{A}}_{\Lambda \hspace{0.01in}\text{ev}}^{(1)}$ is evanescent and
possibly nonlocal.

Now we take the divergent part of equation (\ref{ana2}). Decompose $\tilde{%
\Gamma}_{\Lambda }^{(1)}$ as the sum of its divergent part $\tilde{\Gamma}%
_{\Lambda \hspace{0.01in}\text{div}}^{(1)}$ and its convergent part $\tilde{%
\Gamma}_{\Lambda \hspace{0.01in}\text{conv}}^{(1)}$. Recalling that the
antiparentheses of convergent functionals are convergent, we obtain that $(%
\tilde{S}_{\Lambda },\tilde{\Gamma}_{\Lambda \hspace{0.01in}\text{conv}%
}^{(1)})$ is convergent. Properties (\ref{loco}) apply to $(\tilde{S}%
_{\Lambda },\tilde{\Gamma}_{\Lambda \hspace{0.01in}\text{div}}^{(1)})$, so
we have the identity 
\begin{equation}
(\tilde{S}_{\Lambda },\tilde{\Gamma}_{\Lambda \hspace{0.01in}\text{div}%
}^{(1)})=\frac{1}{2}\mathcal{\tilde{A}}_{\Lambda \hspace{0.01in}\text{divev}%
}^{(1)}.  \label{noveva}
\end{equation}%
Now, formula (\ref{con2}) tells us that $\tilde{\Gamma}_{\Lambda \hspace{%
0.01in}\text{div}}^{(1)}$ is just a functional of $\tilde{g}\tilde{A}$.
Therefore, its antiparenthesis with $\tilde{S}_{\Lambda }$ is only sensitive
to $\tilde{S}_{K}$ and the $K$-dependent contributions to $\tilde{S}%
_{\Lambda \hspace{0.01in}\text{ev}}$, which we denote by $\tilde{S}_{\Lambda
K\hspace{0.01in}\text{ev}}$. Moreover, we can further decompose $\tilde{%
\Gamma}_{\Lambda \hspace{0.01in}\text{div}}^{(1)}$ as the sum of a
nonevanescent divergent part $\tilde{\Gamma}_{\Lambda \hspace{0.01in}\text{%
nev\hspace{0.01in}div}}^{(1)}$ and a divergent evanescence $\tilde{\Gamma}%
_{\Lambda \hspace{0.01in}\text{divev}}^{(1)}$. So doing, we find 
\begin{equation}
(\tilde{S}_{K}+\tilde{S}_{\Lambda K\hspace{0.01in}\text{ev}},\tilde{\Gamma}%
_{\Lambda \hspace{0.01in}\text{nev\hspace{0.01in}div}}^{(1)}+\tilde{\Gamma}%
_{\Lambda \hspace{0.01in}\text{divev}}^{(1)})=\frac{1}{2}\mathcal{\tilde{A}}%
_{\Lambda \hspace{0.01in}\text{divev}}^{(1)}.  \label{nov3}
\end{equation}%
At this point, taking the nonevanescent divergent part of this equation,\ we
obtain 
\begin{equation*}
(\tilde{S}_{K},\tilde{\Gamma}_{\Lambda \hspace{0.01in}\text{nev\hspace{0.01in%
}div}}^{(1)})=0,
\end{equation*}%
which just states that $\tilde{\Gamma}_{\Lambda \hspace{0.01in}\text{nev%
\hspace{0.01in}div}}^{(1)}$ is gauge invariant. Going back to the nontilde
parametrization, we have $\tilde{\Gamma}_{\Lambda \hspace{0.01in}\text{nev%
\hspace{0.01in}div}}^{(1)}(\tilde{g}\tilde{A})=\Gamma _{\Lambda \hspace{%
0.01in}\text{nev\hspace{0.01in}div}}^{(1)}(gA)$. By power counting, $\Gamma
_{\Lambda \hspace{0.01in}\text{nev\hspace{0.01in}div}}^{(1)}$ can only be a
linear combination of the invariants $F_{\mu \nu }^{a_{i}}F^{a_{i}\hspace{%
0.01in}\mu \nu }$, and can be subtracted by redefining the parameters $\zeta
_{i}$. The rest, $\Gamma _{\Lambda \hspace{0.01in}\text{divev}}^{(1)}$, can
be subtracted by redefining the parameters $\eta $ of $S_{\text{ev}}$. The
renormalized action $\hat{S}_{\Lambda }$ is obtained by making the
replacements 
\begin{equation}
\zeta _{i}\rightarrow \zeta _{i}+\frac{f_{i}}{\varepsilon }g^{2},\text{%
\qquad }\eta \rightarrow \eta +\frac{f^{\prime }}{\varepsilon }g^{2},
\label{rena}
\end{equation}%
in $S_{\Lambda }$, where $f_{i}$, $f^{\prime }$ are calculable numerical
coefficients. Since $S_{\Lambda }$ is linear in $\zeta $ and $\eta $, we
have 
\begin{equation}
\hat{S}_{\Lambda }=\tilde{S}_{\Lambda }-\tilde{\Gamma}_{\Lambda \hspace{%
0.01in}\text{div}}^{(1)}.  \label{hatta}
\end{equation}%
Moreover, using (\ref{noveva}) and $(\tilde{\Gamma}_{\Lambda \hspace{0.01in}%
\text{div}}^{(1)},\tilde{\Gamma}_{\Lambda \hspace{0.01in}\text{div}}^{(1)})=0
$ we find 
\begin{equation}
(\hat{S}_{\Lambda },\hat{S}_{\Lambda })=(\tilde{S}_{\Lambda },\tilde{S}%
_{\Lambda })-\mathcal{\tilde{A}}_{\Lambda \hspace{0.01in}\text{divev}}^{(1)}.
\label{hattadiv}
\end{equation}

The generating functional $\hat{\Gamma}_{\Lambda }$ defined by $\hat{S}%
_{\Lambda }$ is convergent to all orders, because formula (\ref{counter})
ensures that no divergences can appear beyond one loop. Finally, $\hat{\Gamma%
}_{\Lambda }$ and the anomaly $\mathcal{\hat{A}}_{\Lambda }=(\hat{\Gamma}%
_{\Lambda },\hat{\Gamma}_{\Lambda })$ are obtained by making the
replacements (\ref{rena}) inside $\tilde{\Gamma}_{\Lambda }$ and $\mathcal{%
\tilde{A}}_{\Lambda }=(\tilde{\Gamma}_{\Lambda },\tilde{\Gamma}_{\Lambda })$%
, respectively. Clearly, $\mathcal{\hat{A}}_{\Lambda }$ is convergent,
because $\hat{\Gamma}_{\Lambda }$ is convergent, and because the
antiparentheses of convergent functionals are convergent.

\section{One-loop anomalies}

\setcounter{equation}{0}

In this section we study the one-loop anomalies, and relate those of the
final theory, which are trivial by assumption, to those of the
higher-derivative theory, which turn out to be trivial as a consequence.

We begin with the one-loop contributions $\mathcal{\hat{A}}_{\Lambda }^{(1)}$
and $\mathcal{\tilde{A}}_{\Lambda }^{(1)}$ to $\mathcal{\hat{A}}_{\Lambda }$
and $\mathcal{\tilde{A}}_{\Lambda }$. First, we observe that 
\begin{equation*}
\mathcal{\hat{A}}_{\Lambda }=\langle (\hat{S}_{\Lambda },\hat{S}_{\Lambda
})\rangle _{\hat{S}_{\Lambda }}=\langle (\hat{S}_{\Lambda },\hat{S}_{\Lambda
})\rangle _{\tilde{S}_{\Lambda }-\tilde{\Gamma}_{\Lambda \hspace{0.01in}%
\text{div}}^{(1)}}=\langle (\hat{S}_{\Lambda },\hat{S}_{\Lambda })\rangle _{%
\tilde{S}_{\Lambda }}+\mathcal{O}(\hbar ^{2}).
\end{equation*}%
Indeed, the correction $\tilde{\Gamma}_{\Lambda \hspace{0.01in}\text{div}%
}^{(1)}$ to the action provides $\mathcal{O}(\hbar )$ vertices. If we use
those vertices in one-particle irreducible diagrams together with vertices
of $(\hat{S}_{\Lambda },\hat{S}_{\Lambda })$, we must close at least one
loop, which gives $\mathcal{O}(\hbar ^{2})$ contributions. Using (\ref%
{hattadiv}), we have 
\begin{equation*}
\mathcal{\hat{A}}_{\Lambda }=\langle (\tilde{S}_{\Lambda },\tilde{S}%
_{\Lambda })\rangle _{\tilde{S}_{\Lambda }}-\mathcal{\tilde{A}}_{\Lambda 
\hspace{0.01in}\text{divev}}^{(1)}+\mathcal{O}(\hbar ^{2})=\mathcal{\tilde{A}%
}_{\Lambda }-\mathcal{\tilde{A}}_{\Lambda \hspace{0.01in}\text{divev}}^{(1)}+%
\mathcal{O}(\hbar ^{2}),
\end{equation*}%
thus (\ref{adivev}) gives 
\begin{equation}
\mathcal{\hat{A}}_{\Lambda }^{(1)}=\mathcal{\tilde{A}}_{\Lambda \hspace{%
0.01in}\text{nev}}^{(1)}+\mathcal{\tilde{A}}_{\Lambda \hspace{0.01in}\text{ev%
}}^{(1)}.  \label{adivevhat}
\end{equation}%
As a check, recall that $\mathcal{\hat{A}}_{\Lambda }$ is convergent, so the
divergent evanescences $\mathcal{\tilde{A}}_{\Lambda \hspace{0.01in}\text{%
divev}}^{(1)}$ must disappear from $\mathcal{\hat{A}}_{\Lambda }^{(1)}$.

We know that $\mathcal{\tilde{A}}_{\Lambda \hspace{0.01in}\text{nev}}^{(1)}$
is the integral of a local function of dimension 5 and ghost number 1.
Recalling that a factor $\tilde{g}$ is attached to every external leg, we
have 
\begin{equation}
\mathcal{\tilde{A}}_{\Lambda \hspace{0.01in}\text{nev}}^{(1)}=\int \mathrm{d}%
^{D}x\ \tilde{g}\tilde{C}^{a}\mathcal{\tilde{A}}^{a}(\tilde{g}\tilde{\Phi},%
\tilde{g}\tilde{K}),  \label{struct5}
\end{equation}%
where $\mathcal{\tilde{A}}^{a}$ are local functions of ghost number zero and
dimension 4. However, $\mathcal{\tilde{A}}_{\Lambda \hspace{0.01in}\text{nev}%
}^{(1)}$ cannot depend on the sources $\tilde{K}$ and the matter fields $%
\tilde{\psi}$, because the products $\tilde{g}\tilde{K}$ and $\tilde{g}%
\tilde{\psi}$ have dimensions greater than 4.

Working out $(\tilde{S}_{\Lambda },\tilde{S}_{\Lambda })$ in detail, it is
easy to check that it does not depend on $\tilde{B}^{a_{i}}$ and depends on $%
\tilde{K}^{\mu a_{i}}$ and $\widetilde{\bar{C}}^{a_{i}}$ only through the
combinations $\tilde{K}^{\mu a_{i}}+\tilde{Q}(\Box )\partial ^{\mu }%
\widetilde{\bar{C}}^{a_{i}}$. Therefore, the same must be true of $\mathcal{%
\tilde{A}}_{\Lambda }^{(1)}$, which means that $\mathcal{\tilde{A}}_{\Lambda 
\hspace{0.01in}\text{nev}}^{(1)}$ cannot depend on either $\widetilde{\bar{C}%
}$ or $\tilde{B}$. Then the functions $\mathcal{\tilde{A}}^{a}$ cannot even
contain ghosts. Summarizing, we can write 
\begin{equation}
\mathcal{\tilde{A}}_{\Lambda \hspace{0.01in}\text{nev}}^{(1)}=\int \mathrm{d}%
^{D}x\ \tilde{g}\tilde{C}^{a}\mathcal{\tilde{A}}^{a}(\tilde{g}\tilde{A}).
\label{line}
\end{equation}

Recall that the antiparentheses satisfy the identity $(X,(X,X))=0$ for any
functional $X$. Taking $X=\hat{\Gamma}_{\Lambda }$, we obtain 
\begin{equation}
(\hat{\Gamma}_{\Lambda },\mathcal{\hat{A}}_{\Lambda })=0,  \label{acca}
\end{equation}%
which are the Wess-Zumino consistency conditions \cite{wesszumino}, written
using the Batalin-Vilkovisky formalism. In particular, at one loop we have 
\begin{equation}
(\tilde{S}_{\Lambda },\mathcal{\hat{A}}_{\Lambda }^{(1)})=-(\hat{\Gamma}%
_{\Lambda }^{(1)},(\tilde{S}_{\Lambda },\tilde{S}_{\Lambda })).  \label{rhs}
\end{equation}%
In section 2 we have proved that the antiparenthesis of an evanescent
functional with a convergent functional is evanescent. Thus, 
\begin{equation*}
(\hat{\Gamma}_{\Lambda }^{(1)},(\tilde{S}_{\Lambda },\tilde{S}_{\Lambda }))=%
\text{ev}=\mathcal{O}(\varepsilon ).
\end{equation*}%
For the same reason, $(\tilde{S}_{\Lambda },\mathcal{\tilde{A}}_{\Lambda 
\hspace{0.01in}\text{ev}}^{(1)})$ and $(\tilde{S}_{\Lambda K\hspace{0.01in}%
\text{ev}},\mathcal{\tilde{A}}_{\Lambda \hspace{0.01in}\text{nev}}^{(1)})$
are evanescent. Using these facts, together with (\ref{adivevhat}) and (\ref%
{line}), formula (\ref{rhs}) gives 
\begin{equation*}
\text{ev}=(\tilde{S}_{\Lambda },\mathcal{\tilde{A}}_{\Lambda \hspace{0.01in}%
\text{nev}}^{(1)}+\mathcal{\tilde{A}}_{\Lambda \hspace{0.01in}\text{ev}%
}^{(1)})=(\tilde{S}_{\Lambda },\mathcal{\tilde{A}}_{\Lambda \hspace{0.01in}%
\text{nev}}^{(1)})+\text{ev}=(\tilde{S}_{K}+\tilde{S}_{\Lambda K\hspace{%
0.01in}\text{ev}},\mathcal{\tilde{A}}_{\Lambda \hspace{0.01in}\text{nev}%
}^{(1)})+\text{ev}=(\tilde{S}_{K},\mathcal{\tilde{A}}_{\Lambda \hspace{0.01in%
}\text{nev}}^{(1)})+\text{ev.}
\end{equation*}%
At this point, we take the nonevanescent part of both sides and note that
relations (\ref{loco}) apply to $(\tilde{S}_{K},\mathcal{\tilde{A}}_{\Lambda 
\hspace{0.01in}\text{nev}}^{(1)})$, because, thanks to (\ref{line}), no
products of more fermion bilinears are involved in these antiparentheses. We
find 
\begin{equation}
(\tilde{S}_{K},\mathcal{\tilde{A}}_{\Lambda \hspace{0.01in}\text{nev}%
}^{(1)})=0.  \label{aqua}
\end{equation}

Now, $\mathcal{\tilde{A}}_{\Lambda \hspace{0.01in}\text{nev}}^{(1)}$ is the
(potential) one-loop anomaly of the higher-derivative regularized theory $%
\tilde{S}_{\Lambda }$, defined keeping $\Lambda $ fixed. The final theory is
instead obtained taking the DHD limit. We must relate $\mathcal{\tilde{A}}%
_{\Lambda \hspace{0.01in}\text{nev}}^{(1)}$ to the potential one-loop
anomaly $\mathcal{A}_{\hspace{0.01in}f\hspace{0.01in}\text{nev}}^{(1)}$ of
the final theory. Indeed, we are assuming that $\mathcal{A}_{\hspace{0.01in}f%
\hspace{0.01in}\text{nev}}^{(1)}$ is trivial (the final theory cannot have
gauge anomalies at one loop), but we have no information of this type as
regards $\mathcal{\tilde{A}}_{\Lambda \hspace{0.01in}\text{nev}}^{(1)}$.

We know how $\mathcal{\tilde{A}}_{\Lambda \hspace{0.01in}\text{nev}}^{(1)}$
depends on $\tilde{g}$. The other dimensionful parameters of $\tilde{S}%
_{\Lambda }$ (such as $\tilde{\zeta}_{i}$ and $\tilde{\xi}_{i}$), as well as
the powers of $\Lambda $ multiplying various terms (such as $\widetilde{\bar{%
\psi}}_{L}^{I}\imath \widetilde{\slashed{D}}\tilde{\psi}_{L}^{I}$), have
dimensions greater than 4. They cannot contribute to $\mathcal{\tilde{A}}%
_{\Lambda \hspace{0.01in}\text{nev}}^{(1)}$, because the local functions $%
\mathcal{\tilde{A}}^{a}$ are polynomial in them and have dimension 4. Thus, $%
\mathcal{\tilde{A}}_{\Lambda \hspace{0.01in}\text{nev}}^{(1)}$ can only
depend on $\tilde{g}\tilde{C}$, $\tilde{g}\tilde{A}$, $\tilde{r}_{i}$, $%
\lambda ^{\prime }$, $\xi ^{\prime }$, $\eta _{1i}$ and $\eta _{2i}$. Using (%
\ref{line}), switching to nontilde variables, and recalling that $\tilde{g}%
\tilde{A}=gA$, $\tilde{g}\tilde{C}=gC$, we obtain that $\mathcal{A}_{\Lambda 
\hspace{0.01in}\text{nev}}^{(1)}$ is $\Lambda $ independent. Now we show
that actually $\mathcal{A}_{\Lambda \hspace{0.01in}\text{nev}}^{(1)}$
coincides with the one-loop anomaly $\mathcal{A}_{\hspace{0.01in}f\hspace{%
0.01in}\text{nev}}^{(1)}$ of the final theory.

To prove this fact, we need to take $\Lambda $ to infinity and study the
DHD\ limit at one loop. A more comprehensive study of the DHD limit will be
carried out later. The terms that are divergent in this limit are denoted by
``Ddiv'', to distinguish them from the divergences considered so far, which
strictly speaking were ``$\varepsilon $div''. Recall that, according to the
definition of DHD limit, the $\Lambda $-divergent parts cannot contain
analytic $\varepsilon $ evanescences, but can contain formal $\varepsilon $
evanescences.

Consider $\mathcal{\hat{A}}_{\Lambda }=(\hat{\Gamma}_{\Lambda },\hat{\Gamma}%
_{\Lambda })$ and take the one-loop DHD-divergent part of this equation.
Using (\ref{adivevhat}) and recalling that $\mathcal{A}_{\Lambda \hspace{%
0.01in}\text{nev}}^{(1)}$ is $\Lambda $ independent, we get 
\begin{eqnarray}
\frac{1}{2}\left. \mathcal{A}_{\Lambda \hspace{0.01in}\text{ev\hspace{0.01in}%
}}^{(1)}\right\vert _{\text{Ddiv}} &=&\left. (S_{\Lambda },\hat{\Gamma}%
_{\Lambda }^{(1)})\right\vert _{\text{Ddiv}}=\left. (S_{\Lambda },\hat{\Gamma%
}_{\Lambda \hspace{0.01in}\text{Ddiv}}^{(1)})\right\vert _{\text{Ddiv}} 
\notag \\
&=&\left. (S_{\Lambda }-S_{r},\hat{\Gamma}_{\Lambda \hspace{0.01in}\text{Ddiv%
}}^{(1)})\right\vert _{\text{Ddiv}}+\left. (S_{r},\hat{\Gamma}_{\Lambda 
\hspace{0.01in}\text{Ddiv}}^{(1)})\right\vert _{\text{Ddiv}}=(S_{r},\hat{%
\Gamma}_{\Lambda \hspace{0.01in}\text{Ddiv}}^{(1)}),  \label{bingo}
\end{eqnarray}%
where $\hat{\Gamma}_{\Lambda \hspace{0.01in}\text{Ddiv}}^{(1)}$ is the
one-loop DHD-divergent part of $\hat{\Gamma}_{\Lambda }$. In the last step
we have dropped the contribution involving $(S_{\Lambda }-S_{r},\hat{\Gamma}%
_{\Lambda \hspace{0.01in}\text{Ddiv}}^{(1)})$, since this quantity vanishes
in the limit $\Lambda \rightarrow \infty $. The reason is that, by formulas (%
\ref{exta}) and (\ref{sl}), the difference $S_{\Lambda }-S_{r}$ is made of $%
\mathcal{O}(1/\Lambda ^{6})$ terms, and the powerlike $\Lambda $ divergences
contained in $\hat{\Gamma}_{\Lambda \hspace{0.01in}\text{Ddiv}}^{(1)}$
cannot exceed $\Lambda ^{4}$. Actually, this is one of the reasons why we
have chosen the particular higher-derivative structure of the theory $%
S_{\Lambda }$. Moreover, to make the last step of (\ref{bingo}) we have
applied (\ref{loco}) to $(S_{r},\hat{\Gamma}_{\Lambda \hspace{0.01in}\text{%
Ddiv}}^{(1)})$. Because of the analysis of section 3, the $\Lambda $
divergences of $\hat{\Gamma}_{\Lambda \hspace{0.01in}\text{Ddiv}}^{(1)}$ can
be of two types, with respect to the limit $\varepsilon \rightarrow 0$:
nonevanescent or formally evanescent. Thanks to (\ref{loco}), the
antiparentheses with $S_{r}$ also give nonevanescent or formally evanescent
contributions, wherefrom the last equality of (\ref{bingo}) follows.

Subtracting the $\Lambda $ divergences $\hat{\Gamma}_{\Lambda \hspace{0.01in}%
\text{Ddiv}}^{(1)}$ from $\hat{S}_{\Lambda }$, we can define the one-loop
renormalized action $\hat{S}_{f\hspace{0.01in}\text{ren}}$ of the final
theory, which reads 
\begin{equation*}
\hat{S}_{f\hspace{0.01in}\text{ren}}=\hat{S}_{\Lambda }-\hat{\Gamma}%
_{\Lambda \hspace{0.01in}\text{Ddiv}}^{(1)}+\mathcal{O}(\hbar ^{2}).
\end{equation*}%
For the moment we do not need to specify the $\mathcal{O}(\hbar ^{2})$ terms
of this subtraction (but later we will have to be precise about them). The
anomaly of the final theory is 
\begin{equation*}
\mathcal{A}_{f}=\langle (\hat{S}_{f\hspace{0.01in}\text{ren}},\hat{S}_{f%
\hspace{0.01in}\text{ren}})\rangle _{\hat{S}_{f\hspace{0.01in}\text{ren}}},
\end{equation*}%
and its one-loop nonevanescent part is the quantity $\mathcal{A}_{\hspace{%
0.01in}f\hspace{0.01in}\text{nev}}^{(1)}$ we want, where the subscript
\textquotedblleft nev\textquotedblright\ close to the subscript
\textquotedblleft $f$\textquotedblright\ denotes the contributions that do
not vanish in the DHD\ limit. We have 
\begin{eqnarray}
\mathcal{A}_{f} &=&\langle (\hat{S}_{\Lambda }-\hat{\Gamma}_{\Lambda \hspace{%
0.01in}\text{Ddiv}}^{(1)},\hat{S}_{\Lambda }-\hat{\Gamma}_{\Lambda \hspace{%
0.01in}\text{Ddiv}}^{(1)})\rangle _{\hat{S}_{\Lambda }-\hat{\Gamma}_{\Lambda 
\hspace{0.01in}\text{Ddiv}}^{(1)}}+\mathcal{O}(\hbar ^{2})=\mathcal{\hat{A}}%
_{\Lambda }-2(S_{\Lambda },\hat{\Gamma}_{\Lambda \hspace{0.01in}\text{Ddiv}%
}^{(1)})+\mathcal{O}(\hbar ^{2})  \notag \\
&=&(S_{\Lambda },S_{\Lambda })+\mathcal{A}_{\Lambda \hspace{0.01in}\text{nev}%
}^{(1)}+\mathcal{A}_{\Lambda \hspace{0.01in}\text{ev}}^{(1)}-2(S_{r},\hat{%
\Gamma}_{\Lambda \hspace{0.01in}\text{Ddiv}}^{(1)})-2(S_{\Lambda }-S_{r},%
\hat{\Gamma}_{\Lambda \hspace{0.01in}\text{Ddiv}}^{(1)})+\mathcal{O}(\hbar
^{2}).  \label{7}
\end{eqnarray}%
In these manipulations we have used the formula 
\begin{equation*}
\mathcal{\hat{A}}_{\Lambda }=\langle (\hat{S}_{\Lambda },\hat{S}_{\Lambda
})\rangle _{\hat{S}_{\Lambda }}=\langle (\hat{S}_{\Lambda },\hat{S}_{\Lambda
})\rangle _{\hat{S}_{\Lambda }-\hat{\Gamma}_{\Lambda \hspace{0.01in}\text{%
Ddiv}}^{(1)}}+\mathcal{O}(\hbar ^{2}),
\end{equation*}%
which holds because at one loop the vertices of $\hat{\Gamma}_{\Lambda 
\hspace{0.01in}\text{Ddiv}}^{(1)}$, which are already $\mathcal{O}(\hbar )$,
cannot contribute to one-particle irreducible diagrams containing one
insertion of $(\hat{S}_{\Lambda },\hat{S}_{\Lambda })$.

At one loop, using (\ref{bingo}), we obtain 
\begin{equation}
\mathcal{A}_{f}^{(1)}=\mathcal{A}_{\Lambda \hspace{0.01in}\text{nev}}^{(1)}+%
\mathcal{A}_{\Lambda \hspace{0.01in}\text{ev}}^{(1)}-\left. \mathcal{A}%
_{\Lambda \hspace{0.01in}\text{ev\hspace{0.01in}}}^{(1)}\right\vert _{\text{%
Ddiv}}-2(S_{\Lambda }-S_{r},\hat{\Gamma}_{\Lambda \hspace{0.01in}\text{Ddiv}%
}^{(1)}).  \label{8}
\end{equation}%
We are ready to take the DHD limit. Recall that $(S_{\Lambda }-S_{r},\hat{%
\Gamma}_{\Lambda \hspace{0.01in}\text{Ddiv}}^{(1)})$ tends to zero for $%
\Lambda \rightarrow \infty $, while $\mathcal{A}_{\Lambda \hspace{0.01in}%
\text{nev}}^{(1)}$ does not change. On the other hand, $\mathcal{A}_{\Lambda 
\hspace{0.01in}\text{ev}}^{(1)}$ and its $\Lambda $-divergent part do not
separately tend to zero, because they can contain (local) terms that are
formally $\varepsilon $ evanescent and $\Lambda $ divergent. However, those
terms are precisely $\left. \mathcal{A}_{\Lambda \hspace{0.01in}\text{ev%
\hspace{0.01in}}}^{(1)}\right\vert _{\text{Ddiv}}$. Therefore, they
disappear in the difference $\mathcal{A}_{\Lambda \hspace{0.01in}\text{ev}%
}^{(1)}-\left. \mathcal{A}_{\Lambda \hspace{0.01in}\text{ev\hspace{0.01in}}%
}^{(1)}\right\vert _{\text{Ddiv}}$. Finally, using (\ref{line}), we get 
\begin{equation}
\mathcal{A}_{\hspace{0.01in}f\hspace{0.01in}\text{nev}}^{(1)}=\mathcal{A}%
_{\Lambda \hspace{0.01in}\text{nev}}^{(1)}=\int \mathrm{d}^{D}x\ gC^{a}%
\mathcal{A}^{a}(gA),  \label{line2}
\end{equation}%
as we wanted.

Let us write the most general structure of the functions $\mathcal{A}%
^{a}(gA) $. We know that they have dimension 4 and are sums of terms of the
form $g^{p}\partial ^{k}A^{p}$. Power counting gives $k+p\leq 4$, hence we
have 
\begin{equation*}
\mathcal{A}^{a}\sim g^{2}\partial ^{2}A^{2}+g^{3}\partial A^{3}+g^{4}A^{4},
\end{equation*}%
plus the terms obtained from these by suppressing some $gA$'s or some
derivatives. Now it remains to collect all pieces of information found so
far and solve (\ref{aqua}). We call condition (\ref{aqua}) a \textit{little}
cohomological problem, because it involves a structure (\ref{line}) that
contains a finite number of terms, in our case just a few, and its solution
can be worked out directly. We recall the solution without proof, because
the proof is well-known and not necessary for the other derivations of this
paper. The solution can be split into the sum of trivial and nontrivial
contributions. Trivial contributions are those of the form $(S_{K},\chi )$,
where $\chi =\chi (gA)$ is a local functional of the gauge fields $A$, equal
to the integral of a local function of dimension 4 and ghost number 0, and
having a $g$ structure corresponding to the one-loop sector of formula (\ref%
{liable2}). In the tilde parametrization, we write $\chi $ as $\tilde{\chi}(%
\tilde{g}\tilde{A})$. The only nontrivial contributions to $\mathcal{A}_{f%
\hspace{0.01in}\text{nev}}^{(1)}$ are proportional to the famous Bardeen
formula \cite{bardeen}. In appendix A, the coefficient of the Bardeen term
is calculated using our regularization technique. In the end, we have 
\begin{equation}
\mathcal{A}_{f\hspace{0.01in}\text{nev}}^{(1)}=\mathcal{A}_{\Lambda \hspace{%
0.01in}\text{nev}}^{(1)}=-\frac{\imath g^{3}}{12\pi ^{2}}\int \mathrm{d}%
^{D}x\ \varepsilon ^{\mu \nu \rho \sigma }\mathrm{Tr}\left[ \partial _{\mu
}C\left( A_{\nu }\partial _{\rho }A_{\sigma }+\frac{g}{2}A_{\nu }A_{\rho
}A_{\sigma }\right) \right] +(S_{K},\chi ),  \label{a1loop}
\end{equation}%
where $C=C^{a}T^{a}$, $A_{\mu }=A_{\mu }^{a}T^{a}$, the Bardeen term being
the integral on the right-hand side.

One-loop gauge anomalies vanish when the trace appearing in (\ref{a1loop})
vanishes. Typically, the cancellation is possible when the gauge group is a
product group and the theory contains various types of fermionic fields in
suitable representations, as in the standard model.

Now we go back to the higher-derivative theory (the DHD limit being
completed in section 7), precisely to the classical action $\hat{S}_{\Lambda
}$ of formula (\ref{hatta}). The trivial contributions $(S_{K},\chi )$ to
anomalies can be canceled out by redefining the action as 
\begin{equation}
\hat{S}_{\Lambda }^{\prime }(\Phi ,K)=\hat{S}_{\Lambda }(\Phi ,K)-\frac{1}{2}%
\chi (gA),  \label{suba}
\end{equation}%
because then 
\begin{equation*}
\mathcal{\hat{A}}_{\Lambda }^{\prime }=\langle (\hat{S}_{\Lambda }^{\prime },%
\hat{S}_{\Lambda }^{\prime })\rangle _{\hat{S}_{\Lambda }^{\prime }}=\langle
(\hat{S}_{\Lambda },\hat{S}_{\Lambda })\rangle _{\hat{S}_{\Lambda }}-(\hat{S}%
_{\Lambda },\chi )+\mathcal{O}(\hbar ^{2})=\mathcal{\hat{A}}_{\Lambda
}-(S_{K}+S_{\Lambda K\hspace{0.01in}\text{ev}},\chi )+\mathcal{O}(\hbar
^{2}).
\end{equation*}%
In the last step we used the fact that $\chi $ is $K$ independent. Thus, at
one loop we have 
\begin{equation*}
\mathcal{\hat{A}}_{\Lambda }^{\prime \hspace{0.01in}(1)}=\mathcal{A}%
_{\Lambda \hspace{0.01in}\text{nev}}^{(1)}+\mathcal{A}_{\Lambda \hspace{%
0.01in}\text{ev}}^{(1)}-(S_{K},\chi )+\text{ev},\qquad \mathcal{\hat{A}}%
_{\Lambda \hspace{0.01in}\text{nev}}^{\prime \hspace{0.01in}(1)}=\mathcal{A}%
_{\Lambda \hspace{0.01in}\text{nev}}^{(1)}-(S_{K},\chi ),
\end{equation*}%
which means that when the Bardeen term vanishes $\mathcal{\hat{A}}_{\Lambda 
\hspace{0.01in}\text{nev}}^{\prime \hspace{0.01in}(1)}=0$, $\mathcal{\hat{A}}%
_{\Lambda }^{\prime \hspace{0.01in}(1)}=$ ev.

Finally, observe that the new $\Gamma $ functional $\hat{\Gamma}_{\Lambda
}^{\prime }$ is still convergent to all orders. The reason is that it is
convergent at one loop and the action 
\begin{equation}
\hat{S}_{\Lambda }^{\prime }=S_{\Lambda }-\Gamma _{\Lambda \hspace{0.01in}%
\text{div}}^{(1)}-\frac{1}{2}\chi  \label{sp}
\end{equation}%
has the $g$ structure (\ref{liable2}). Then, using tilde variables, the
counterterms must have the form (\ref{counter}), which however forbids
divergent contributions from two loops onwards. The anomaly functional $%
\mathcal{\hat{A}}_{\Lambda }^{\prime }=(\hat{\Gamma}_{\Lambda }^{\prime },%
\hat{\Gamma}_{\Lambda }^{\prime })$ is also convergent to all orders and has
the $g$ structure (\ref{liable2}).

The next step is to prove the anomaly cancellation to all orders in the
higher-derivative theory. After that, we will have to complete the DHD limit
by renormalizing the $\Lambda $ divergences.

\section{Manifest Adler-Bardeen theorem in the higher-derivative theory}

\setcounter{equation}{0}

In this section we prove that gauge anomalies manifestly cancel to all
orders in the higher-derivative theory $S_{\Lambda }$. We assume that the
final theory has no one-loop anomalies, which, according to the previous
section, implies that the higher-derivative theory shares the same property,
namely $\mathcal{A}_{\Lambda \hspace{0.01in}\text{nev}}^{(1)}=(S_{K},\chi )$%
, $\mathcal{\hat{A}}_{\Lambda \hspace{0.01in}\text{nev}}^{\prime \hspace{%
0.01in}(1)}=0$. Then, the one-loop contribution $\mathcal{\hat{A}}_{\Lambda
}^{\prime \hspace{0.01in}(1)}$ to the anomaly functional $\mathcal{\hat{A}}%
_{\Lambda }^{\prime }$ is evanescent, so we can write 
\begin{equation}
\mathcal{\hat{A}}_{\Lambda }^{\prime }=\mathcal{O}(\varepsilon )+\mathcal{O}%
(\hbar ^{2}).  \label{becau}
\end{equation}%
Here the \textquotedblleft $\mathcal{O}(\varepsilon )$\textquotedblright\
includes the tree-level contribution $(S_{\Lambda },S_{\Lambda })$.

Now we move on to higher orders. We have to study the diagrams with two or
more loops, and one insertion of 
\begin{equation}
\mathcal{E}\equiv (\hat{S}_{\Lambda }^{\prime },\hat{S}_{\Lambda }^{\prime
})=(\tilde{S}_{\Lambda },\tilde{S}_{\Lambda })-\mathcal{\tilde{A}}_{\Lambda 
\hspace{0.01in}\text{nev}}^{(1)}-\mathcal{\tilde{A}}_{\Lambda \hspace{0.01in}%
\text{divev}}^{(1)}-(\tilde{S}_{\Lambda K\hspace{0.01in}\text{ev}},\tilde{%
\chi}),  \label{e}
\end{equation}%
calculated with the action (\ref{sp}). We have switched back to the tilde
parametrization, used (\ref{noveva}), and replaced $(\tilde{S}_{\Lambda },%
\tilde{\chi})$ by $(\tilde{S}_{K\hspace{0.01in}}+\tilde{S}_{\Lambda K\hspace{%
0.01in}\text{ev}},\tilde{\chi})$ and $(\tilde{S}_{K},\tilde{\chi})$ by $%
\mathcal{\tilde{A}}_{\Lambda \hspace{0.01in}\text{nev}}^{(1)}$. Both $%
\mathcal{E}$ and $\mathcal{\hat{A}}_{\Lambda }^{\prime }$ have the structure
(\ref{counter}) and $(\tilde{S}_{\Lambda },\tilde{S}_{\Lambda })$ is
formally evanescent. To fix the notation, let us start from formula (\ref%
{corro}), applied to the $\ell $-loop diagrams containing one $(\tilde{S}%
_{\Lambda },\tilde{S}_{\Lambda })$ insertion. We write them as sums of
contributions of the form 
\begin{equation}
\mathcal{G}_{\mathcal{A}}^{(\ell )}=\int \tilde{\Phi}^{\alpha
_{1}}(k_{1})\cdots \tilde{\Phi}^{\alpha _{n}}(k_{n})\tilde{K}_{\beta
_{1}}(k_{n+1})\cdots \tilde{K}_{\beta _{r}}(k_{n+r})\hspace{0.01in}T_{%
\mathcal{A}\mu _{1}\cdots \mu _{p}\alpha _{1}\cdots \alpha _{n}}^{(\ell
)\beta _{1}\cdots \beta _{r}}G_{\mathcal{A}}^{(\ell )\mu _{1}\cdots \mu
_{p}}(k_{1},\cdots ,k_{n+r}),  \label{gla}
\end{equation}%
where the tensors $T_{\mathcal{A}\mu _{1}\cdots \mu _{p}\alpha _{1}\cdots
\alpha _{n}}^{(\ell )\beta _{1}\cdots \beta _{r}}$ are constant and
evanescent, and the integrations over momenta are understood. We recall that 
$G_{\mathcal{A}}^{(\ell )\mu _{1}\cdots \mu _{p}}(k_{1},\cdots ,k_{n+r})$
are the integrals coming from Feynman diagrams, once all tensors $\eta _{\mu
\nu }$, $\varepsilon _{\mu \nu \rho \sigma }$, $\delta _{\hat{\mu}\hat{\nu}}$%
, the $\gamma $ matrices, the structure constants $f^{abc}$ and the matrices 
$T^{a}$ are moved outside into the structures $T_{\mathcal{A}}^{(\ell )}$.\
We call the divergent parts of $G_{\mathcal{A}}^{(\ell )\mu _{1}\cdots \mu
_{p}}$ \textquotedblleft nontrivial\textquotedblright\ if they are not
killed by the structures $T_{\mathcal{A}}^{(\ell )}$.

Let us first reconsider the case $\ell =1$. It is useful to describe the
right-hand side of (\ref{e}) from the point of view of the integrals $G_{%
\mathcal{A}}^{(1)\mu _{1}\cdots \mu _{p}}$. The divergent parts of $G_{%
\mathcal{A}}^{(1)\mu _{1}\cdots \mu _{p}}$ can be of three types: ($a$)
divergences that are turned into nonevanescent contributions by $T_{\mathcal{%
A}}^{(1)}$, which are subtracted by $\mathcal{\tilde{A}}_{\Lambda \hspace{%
0.01in}\text{nev}}^{(1)}$; ($b$) divergences that remain divergent when $T_{%
\mathcal{A}}^{(1)}$ is applied to them, which are subtracted by $\mathcal{%
\tilde{A}}_{\Lambda \hspace{0.01in}\text{divev}}^{(1)}$; ($c$) divergences
that are turned into evanescences by $T_{\mathcal{A}}^{(1)}$, which can be
subtracted by further, one-loop, local evanescent terms $\tilde{L}_{\text{ev}%
}^{(1)}$ with the $\tilde{g}$ structure (\ref{counter}). We write $\mathcal{E%
}=\mathcal{E}_{1}+\mathcal{E}_{2}$, where%
\begin{equation*}
\mathcal{E}_{1}=(\tilde{S}_{\Lambda },\tilde{S}_{\Lambda })-\mathcal{\tilde{A%
}}_{\Lambda \hspace{0.01in}\text{nev}}^{(1)}-\mathcal{\tilde{A}}_{\Lambda 
\hspace{0.01in}\text{divev}}^{(1)}-\tilde{L}_{\text{ev}}^{(1)},\qquad 
\mathcal{E}_{2}=\tilde{L}_{\text{ev}}^{(1)}-(\tilde{S}_{\Lambda K\hspace{%
0.01in}\text{ev}},\tilde{\chi}).
\end{equation*}%
The subtractions included in $\mathcal{E}_{1}$ cancel all nontrivial
divergences of $G_{\mathcal{A}}^{(1)\mu _{1}\cdots \mu _{p}}$. Instead, $%
\left\langle \mathcal{E}_{2}\right\rangle $ collects the diagrams with one $%
\mathcal{E}_{2}$ insertion. They can also be expressed in the form (\ref{gla}%
) and studied along the same lines. From now on we understand that the
expressions (\ref{gla}) refer to the diagrams with one $(\tilde{S}_{\Lambda
},\tilde{S}_{\Lambda })$ insertion or one $\mathcal{E}_{2}$ insertion.

Each contribution $\mathcal{G}_{\mathcal{A}}^{(1)}$ is then equipped with
counterterms $\mathcal{G}_{\mathcal{A}\text{\hspace{0.01in}counter}}^{(1)}$,
so that the difference 
\begin{equation*}
\mathcal{G}_{\mathcal{A}}^{(1)}-\mathcal{G}_{\mathcal{A}\text{\hspace{0.01in}%
counter}}^{(1)}=\int \tilde{\Phi}^{\alpha _{1}}(k_{1})\cdots \tilde{\Phi}%
^{\alpha _{n}}(k_{n})\tilde{K}_{\beta _{1}}(k_{n+1})\cdots \tilde{K}_{\beta
_{r}}(k_{n+r})\hspace{0.01in}T_{\mathcal{A}\mu _{1}\cdots \mu _{p}\alpha
_{1}\cdots \alpha _{n}}^{(1)\beta _{1}\cdots \beta _{r}}G_{\mathcal{A\hspace{%
0.01in}}\text{subtr}}^{(1)\mu _{1}\cdots \mu _{p}}(k),
\end{equation*}%
involves fully convergent subtracted integrals $G_{\mathcal{A\hspace{0.01in}}%
\text{subtr}}^{(1)\mu _{1}\cdots \mu _{p}}$. Now, the evanescences provided
by $T_{\mathcal{A}}^{(1)}$ cannot simplify any divergences, so the final
result $\mathcal{\hat{A}}_{\Lambda }^{\prime \hspace{0.01in}(1)}$ is
evanescent, in agreement with (\ref{becau}).

At higher loops it is useful to make a similar analysis. We begin with $\ell
=2$. The integrals $G_{\mathcal{A}}^{(2)\mu _{1}\cdots \mu _{p}}$ are
automatically equipped with the counterterms that subtract their nontrivial
subdivergences: first, the action $\hat{S}_{\Lambda }^{\prime }$ is equipped
with its own counterterms and, second, the subtractions contained in $%
\mathcal{E}_{1}$ provide counterterms for the integrals $G_{\mathcal{A}%
}^{(1)\mu _{1}\cdots \mu _{p}}$ associated with $(\tilde{S}_{\Lambda },%
\tilde{S}_{\Lambda })$. Instead, the two-loop contributions of $\mathcal{E}%
_{2}$ do not have subdivergences, because $\mathcal{E}_{2}$ is one-loop.
When we include counterterms for subdivergences, we can identify subtracted
integrals $G_{\mathcal{A}}^{(2)\mu _{1}\cdots \mu _{p}}-G_{\mathcal{A\hspace{%
0.01in}}\text{subdiv}}^{(2)\mu _{1}\cdots \mu _{p}}$ that have local
divergent parts $G_{\mathcal{A\hspace{0.01in}}\text{div}}^{(2)\mu _{1}\cdots
\mu _{p}}$ (by the theorem of locality of counterterms) and possibly
nonlocal finite parts $G_{\mathcal{A\hspace{0.01in}}\text{finite}}^{(2)\mu
_{1}\cdots \mu _{p}}$. When $T_{\mathcal{A}}^{(2)}$ acts on $G_{\mathcal{A%
\hspace{0.01in}}\text{div}}^{(2)\mu _{1}\cdots \mu _{p}}$, it gives local
contributions to $\mathcal{\hat{A}}_{\Lambda }^{\prime \hspace{0.01in}(2)}$,
which can be nonevanescent (due to simplified divergences), evanescent or
still divergent. However, local contributions must have the structure (\ref%
{counter}), which implies that they are zero. Indeed, using the tilde
parametrization, they are polynomial in the dimensionful parameters of $%
\tilde{S}_{\Lambda }$ and carry an overall factor $\tilde{g}^{2}$, which has
dimension 16. We conclude that the overall divergences $G_{\mathcal{A\hspace{%
0.01in}}\text{div}}^{(2)\mu _{1}\cdots \mu _{p}}$ are trivial, because they
are killed by $T_{\mathcal{A}}^{(2)}$. When $T_{\mathcal{A}}^{(2)}$ acts on $%
G_{\mathcal{A\hspace{0.01in}}\text{finite}}^{(2)\mu _{1}\cdots \mu _{p}}$ it
just gives (possibly nonlocal) evanescent contributions to $\mathcal{\hat{A}}%
_{\Lambda }^{\prime \hspace{0.01in}(2)}$. Finally, we have 
\begin{equation}
\mathcal{\hat{A}}_{\Lambda }^{\prime \hspace{0.01in}(2)}=\mathcal{O}%
(\varepsilon )\text{.}  \label{2l}
\end{equation}%
Therefore, formula (\ref{becau}) is promoted to the next order, and we can
write $\mathcal{\hat{A}}_{\Lambda }^{\prime }=\mathcal{O}(\varepsilon )+%
\mathcal{O}(\hbar ^{3})$, where now \textquotedblleft $\mathcal{O}%
(\varepsilon )$\textquotedblright\ includes the evanescent contributions
appearing on the right-hand side of (\ref{2l}).

At this point we can proceed by induction. Assume that for some $\ell
\geqslant 2,$%
\begin{equation}
\mathcal{\hat{A}}_{\Lambda }^{\prime }=\mathcal{O}(\varepsilon )+\mathcal{O}%
(\hbar ^{\ell +1}),  \label{assulo}
\end{equation}%
and that the overall divergent parts $G_{\mathcal{A\hspace{0.01in}}\text{div}%
}^{(L)\mu _{1}\cdots \mu _{p}}$ of the subtracted integrals $G_{\mathcal{A}%
}^{(L)\mu _{1}\cdots \mu _{p}}-G_{\mathcal{A\hspace{0.01in}}\text{subdiv}%
}^{(L)\mu _{1}\cdots \mu _{p}}$ are trivial for $2\leqslant L\leqslant \ell $%
. Denote the contributions of order $\hbar ^{\ell +1}$ to $\mathcal{\hat{A}}%
_{\Lambda }^{\prime }$ with $\mathcal{\hat{A}}_{\Lambda }^{\prime \hspace{%
0.01in}(\ell +1)}$. A diagrammatic analysis similar to the one carried out
above shows that $\mathcal{\hat{A}}_{\Lambda }^{\prime \hspace{0.01in}(\ell
+1)}$ is the sum of a local part $\mathcal{\hat{A}}_{\Lambda \text{\hspace{%
0.01in}loc}}^{\prime \hspace{0.01in}(\ell +1)}=\sum T_{\mathcal{A}}^{(\ell
+1)}G_{\mathcal{A\hspace{0.01in}}\text{div}}^{(\ell +1)}$, plus a possibly
nonlocal evanescent part $\mathcal{\hat{A}}_{\Lambda \hspace{0.01in}\text{ev}%
}^{\prime \hspace{0.01in}(\ell +1)}=\sum T_{\mathcal{A}}^{(\ell +1)}G_{%
\mathcal{A\hspace{0.01in}}\text{finite}}^{(\ell +1)}$. However, $\mathcal{%
\hat{A}}_{\Lambda \text{\hspace{0.01in}loc}}^{\prime \hspace{0.01in}(\ell
+1)}$ must have the structure (\ref{counter}), which means that it vanishes.
In the end, $G_{\mathcal{A\hspace{0.01in}}\text{div}}^{(\ell +1)\mu
_{1}\cdots \mu _{p}}$ are also trivial, and $\mathcal{\hat{A}}_{\Lambda
}^{\prime \hspace{0.01in}(\ell +1)}=\mathcal{\hat{A}}_{\Lambda \hspace{0.01in%
}\text{ev}}^{\prime \hspace{0.01in}(\ell +1)}$. Thus, if the inductive
assumptions hold for some $\ell $, they must also hold with $\ell
\rightarrow \ell +1$ and therefore for $\ell =\infty $. We conclude that the
anomaly is evanescent to all orders: 
\begin{equation}
\mathcal{\hat{A}}_{\Lambda }^{\prime \hspace{0.01in}}=(\hat{\Gamma}_{\Lambda
}^{\prime },\hat{\Gamma}_{\Lambda }^{\prime })=\mathcal{O}(\varepsilon ).
\label{evaeva}
\end{equation}

This result proves that if the final theory is anomaly-free at one loop, the
higher-derivative theory $S_{\Lambda }$ is anomaly-free to all orders. It is
important to stress that the DHD-regularization framework provides the
subtraction scheme where this property is \textit{manifest}: after the
subtraction of $(S_{K},\chi )$ at one loop, no analogous subtractions are
necessary at higher orders.

This is not the final result we want, though. To get there we still need to
take $\Lambda $ to infinity and complete the DHD\ limit.

\section{Manifest Adler-Bardeen theorem in the final theory}

\setcounter{equation}{0}

We are finally ready to study anomaly cancellation to all orders in the
final theory. In this section we study the $\Lambda $ dependence and
complete the DHD limit, according to the rules of subsection 3.1. The
subtraction of $\Lambda $ divergences proceeds relatively smoothly, and
preserves the master equation to all orders up to terms that vanish in the
DHD limit.

Call $S_{n}$ and $\Gamma _{n}$ the action and the $\Gamma $ functional
DHD-renormalized up to (and including) $n$ loops, where $S_{0}=\hat{S}%
_{\Lambda }^{\prime }=\hat{S}_{\Lambda }-\chi /2$ is the action (\ref{sp}).
The action $S_{n}$ must satisfy two inductive assumptions to all orders in $%
\hbar $:

(I) $\Gamma _{n}$ has a regular limit for $\varepsilon \rightarrow 0$ at
fixed $\Lambda $, and

(II) the local functional 
\begin{equation}
(S_{n},S_{n})\equiv \mathcal{E}_{n}  \label{ippo}
\end{equation}%
is \textquotedblleft truly $\varepsilon $-evanescent at fixed $\Lambda $%
\textquotedblright , that is to say a local functional such that $\langle 
\mathcal{E}_{n}\rangle $ tends to zero when $\varepsilon \rightarrow 0$ at
fixed $\Lambda $.

More precisely, $\Gamma _{n}$ is a sum of terms (\ref{survo}) up to $n$
loops (because it is DHD-convergent to that order) and a sum of terms (\ref%
{survi}) from $n+1$ loops onwards. Instead, $\langle \mathcal{E}_{n}\rangle
=(\Gamma _{n},\Gamma _{n})$ contains the terms (\ref{survo}) except $%
\varepsilon ^{0}\Lambda ^{0}$ and $\varepsilon ^{0}/\Lambda $ up to $n$
loops, and the terms (\ref{survi}) except $\varepsilon ^{0}$ from $n+1$
loops onwards. Thanks to (\ref{evaeva}) we know that the inductive
hypotheses are true for $n=0$.

The theorem of locality of counterterms ensures that the $(n+1)$-loop
divergent part $\Gamma _{n\hspace{0.01in}\text{div}}^{(n+1)}$ of $\Gamma
_{n} $ is a local functional. Since $\Gamma _{n}$ has a regular limit for $%
\varepsilon \rightarrow 0$ at fixed $\Lambda $, $\Gamma _{n\hspace{0.01in}%
\text{div}}^{(n+1)}$ contains only divergences in $\Lambda $, not in $%
\varepsilon $. Precisely, we can write 
\begin{equation*}
\Gamma _{n\hspace{0.01in}\text{div}}^{(n+1)}=\Gamma _{n\hspace{0.01in}\text{%
div\hspace{0.01in}nev}}^{(n+1)}+\Gamma _{n\hspace{0.01in}\text{div\hspace{%
0.01in}fev}}^{(n+1)},
\end{equation*}
where $\Gamma _{n\hspace{0.01in}\text{div\hspace{0.01in}nev}}^{(n+1)}$ and $%
\Gamma _{n\hspace{0.01in}\text{div\hspace{0.01in}fev}}^{(n+1)}$ collect the
terms $\varepsilon ^{0}\Lambda $ and $\hat{\delta}\varepsilon ^{0}\Lambda $
of the list (\ref{assu}), respectively.

Now we study the $(n+1)$-loop divergent part of $(\Gamma _{n},\Gamma _{n})$.
We take the $(n+1)$-loop DHD-divergent non-$\varepsilon $-evanescent part of 
\begin{equation}
(\Gamma _{n},\Gamma _{n})=\langle (S_{n},S_{n})\rangle =\langle \mathcal{E}%
_{n}\rangle ,  \label{bullu}
\end{equation}
which means the terms of types $\varepsilon ^{0}\Lambda $ of the list (\ref%
{assu}). Recall that $S_{\Lambda }$ is equal to the action $S_{r}$ of (\ref%
{exta}) plus $\mathcal{O}(1/\Lambda ^{6})$ terms, so $(S_{\Lambda
}-S_{r},\Gamma _{n\hspace{0.01in}\text{div}}^{(n+1)})$ is convergent for $%
\Lambda \rightarrow \infty $. Moreover, $S_{r}$ is equal to $S_{\text{gf}}$,
which by formula (\ref{snev}) is non-$\varepsilon $-evanescent, plus $\int 
\bar{\psi}_{R}^{I}\imath \slashed{\partial}\psi _{R}^{I}$ plus $\varepsilon $%
-evanescent terms. Noting that the divergent part of $\langle \mathcal{E}%
_{n}\rangle $ is just made of terms $\hat{\delta}\varepsilon ^{0}\Lambda $,
we obtain 
\begin{equation}
(S_{\text{gf}},\Gamma _{n\hspace{0.01in}\text{div\hspace{0.01in}nev}%
}^{(n+1)})=0.  \label{questa}
\end{equation}
Deriving (\ref{questa}) from (\ref{bullu}) we have expanded $\Gamma
_{n}=\sum_{k=0}^{\infty }\hbar ^{k}\Gamma _{n}^{(k)}$ in powers of $\hbar $
and dropped all contributions $(\Gamma _{n}^{(k)},\Gamma _{n}^{(n+1-k)})$
with $0<k<n+1$, because they are convergent in the DHD limit. Note that $%
\Gamma _{n}^{(k)}$, $0<k<n+1$, may contain terms $\varepsilon \Lambda $.
Now, the powers of $\Lambda $ can get simplified inside $(\Gamma
_{n}^{(k)},\Gamma _{n}^{(n+1-k)})$. However, $\Gamma _{n}$ is convergent for 
$\varepsilon \rightarrow 0$ and the antiparentheses cannot generate poles,
so the resulting contributions remain negligible in the DHD\ limit. We must
just pay attention not to manipulate the terms $\varepsilon \Lambda $ in
inconvenient ways (see subsection 3.1 for details).

Since the theory is power-counting renormalizable, (\ref{questa}) is another
little cohomological problem, therefore it can be solved directly. Moreover,
it is a purely four-dimensional problem, since all $\varepsilon $-evanescent
terms have been dropped. Its solution is well-known and states that $\Gamma
_{n\hspace{0.01in}\text{div\hspace{0.01in}nev}}^{(n+1)}$ can be reabsorbed
by redefining the parameters of $S_{\text{gf}}$ and making a canonical
transformation inside $S_{\text{gf}}$. Using the nonrenormalization of the $%
B $- and $K_{\bar{C}}$-dependent terms, and power counting, the canonical
transformation is generated by a functional 
\begin{eqnarray}
F_{n}(\Phi ,K^{\prime }) &=&\int \sum_{i}(Z_{nAi}^{1/2}A_{\mu
}^{a_{i}}K^{\prime \hspace{0.01in}\mu
a_{i}}+Z_{nCi}^{1/2}C^{a_{i}}K_{C}^{\prime \hspace{0.01in}%
a_{i}}+Z_{nAi}^{-1/2}\bar{C}^{a_{i}}K_{\bar{C}}^{\prime \hspace{0.01in}%
a_{i}}+Z_{nAi}^{-1/2}B^{a_{i}}K_{B}^{\prime \hspace{0.01in}a_{i}})  \notag \\
&&\qquad +\int \left( \bar{\psi}_{L}^{I}Z_{nJI}^{1/2\ast }K_{\psi }^{\prime 
\hspace{0.01in}J}+\bar{K}_{\psi }^{\prime \hspace{0.01in}I}Z_{nIJ}^{1/2}\psi
_{L}^{J}\right) ,  \label{op1}
\end{eqnarray}%
and the parameter redefinitions read 
\begin{equation}
r_{i}^{\prime }=Z_{ni}r_{i},\qquad \xi _{i}^{\prime }=\xi _{i}Z_{nAi},
\label{op2}
\end{equation}%
where $Z_{nAi}$, $Z_{nCi}$, $Z_{nIJ}$ and $Z_{ni}$ are $\varepsilon $%
-independent $\Lambda $-divergent renormalization constants. The $r_{i}$
redefinitions encode the renormalizations of gauge couplings. Instead, the $%
\xi _{i}$ redefinitions follow from the nonrenormalization of the terms
quadratic in $B$. In the parametrization we are using there are no
redefinitions of $g$ and $\zeta _{i}$.

Making the canonical transformation (\ref{op1}) and the redefinitions (\ref%
{op2}) on $S_{\text{gf}}$ we get 
\begin{equation*}
S_{\text{gf}}\rightarrow S_{\text{gf}}-\Gamma _{n\hspace{0.01in}\text{div%
\hspace{0.01in}nev}}^{(n+1)}+\mathcal{O}(\hbar ^{n+2}).
\end{equation*}%
However, the classical action we have been using is not $S_{\text{gf}}$, and
not even $S_{r}=S_{\text{gf}}+S_{\text{LR}}+S_{\text{ev}}$, but $S_{n}$,
whose classical limit is $S_{\Lambda }$, therefore we must inquire what
happens by making the operations (\ref{op1}) and (\ref{op2})\ on $S_{\Lambda
}$.

Let us begin from $S_{r}$. Since $S_{\text{LR}}$ is nonrenormalized, we must
also make the redefinitions 
\begin{equation}
\varsigma _{IJ}^{\prime }=\varsigma _{IK}Z_{nKJ}^{-1/2}.  \label{op3}
\end{equation}
When we apply (\ref{op1}) and (\ref{op2}) to $S_{\text{ev}}$ we generate new
formally $\varepsilon $-evanescent, $\Lambda $-divergent terms of order $%
\hbar ^{n+1}$, which change $\Gamma _{n\hspace{0.01in}\text{div\hspace{0.01in%
}fev}}^{(n+1)}$ into some new $\Gamma _{n\hspace{0.01in}\text{div\hspace{%
0.01in}fev}}^{\prime \hspace{0.01in}(n+1)}$, plus $\mathcal{O}(\hbar ^{n+2})$%
. The divergences $\Gamma _{n\hspace{0.01in}\text{div\hspace{0.01in}fev}%
}^{\prime \hspace{0.01in}(n+1)}$ are not constrained by gauge invariance,
but just locality and power counting. They can be subtracted redefining the
parameters $\eta $ of $S_{\text{ev}}$, since $S_{\text{ev}}$ was added
precisely for this purpose.

We denote the operations that subtract $\Gamma _{n\hspace{0.01in}\text{div}%
}^{(n+1)}$ with $T_{n}$. They include the canonical transformation (\ref{op1}%
), the redefinitions (\ref{op2}) and (\ref{op3}), and the $\eta $
redefinitions that subtract $\Gamma _{n\hspace{0.01in}\text{div\hspace{0.01in%
}fev}}^{\prime \hspace{0.01in}(n+1)}$. Note that $T_{n}=1+\mathcal{O}(\hbar
^{n+1})$. We have 
\begin{equation*}
T_{n}S_{r}=S_{r}-\Gamma _{n\hspace{0.01in}\text{div}}^{(n+1)}+\mathcal{O}%
(\hbar ^{n+2}).
\end{equation*}

It remains to check what happens when the operations $T_{n}$ act on $%
S_{\Lambda }$. Observe that, since no $\varepsilon $ divergences are around,
the operations $T_{n}$ are independent of $\varepsilon $ and divergent in $%
\Lambda $. However, the difference $S_{\Lambda }-S_{r}$ is of order $%
1/\Lambda ^{6}$ and the operations $T_{n}$ cannot contain powers of $\Lambda 
$ greater than 4. Thus, $(T_{n}-1)(S_{\Lambda }-S_{r})$ vanishes in the DHD
limit. Call $S_{n+1}$ the action obtained by applying $T_{n}$ on $S_{n}$. We
have 
\begin{eqnarray}
S_{n+1}=T_{n}S_{n} &=&S_{n}+(T_{n}-1)S_{\Lambda }+\mathcal{O}(\hbar
^{n+2})=S_{n}+(T_{n}-1)S_{r}+(T_{n}-1)(S_{\Lambda }-S_{r})+\mathcal{O}(\hbar
^{n+2})  \notag \\
&=&S_{n}-\Gamma _{n\hspace{0.01in}\text{div}}^{(n+1)}+(T_{n}-1)(S_{\Lambda
}-S_{r})+\mathcal{O}(\hbar ^{n+2}).  \label{tells}
\end{eqnarray}%
This formula tells us that the operations $T_{n}$ do renormalize the
divergences due to $S_{n}$ in the DHD limit. Therefore, $S_{n+1}$ is the $%
(n+1)$-loop DHD-renormalized action, namely it gives a generating functional 
$\Gamma _{n+1}$ that is convergent up to (and including) $n+1$ loops in the
DHD limit.

Moreover, since the canonical transformations generated by (\ref{op1}) act
multiplicatively on fields and sources, the operations $T_{n}$ act on the $%
\Gamma $ functional precisely as they act on the action. Therefore, $\Gamma
_{n+1}=T_{n}\Gamma _{n}$. Since the operations $T_{n}$ are $\varepsilon $%
-independent, we conclude that $\Gamma _{n+1}$ is regular when $\varepsilon
\rightarrow 0$ at fixed $\Lambda $, to all orders in $\hbar $, which
promotes the inductive assumption (I) to $n+1$ loops.

Finally, the operations $T_{n}$ preserve the antiparentheses. Applying them
to (\ref{ippo}) we also obtain 
\begin{equation*}
(S_{n+1},S_{n+1})=T_{n}\mathcal{E}_{n}.
\end{equation*}%
Now, taking the average of this equation we get 
\begin{equation*}
T_{n}\langle \mathcal{E}_{n}\rangle _{n}=T_{n}(\Gamma _{n},\Gamma
_{n})=(\Gamma _{n+1},\Gamma _{n+1})=\langle (S_{n+1},S_{n+1})\rangle
_{n+1}=\langle T_{n}\mathcal{E}_{n}\rangle _{n+1},
\end{equation*}%
where $\langle \mathcal{\cdots }\rangle _{k}$ means that the average is
calculated with the action $S_{k}$. If we take the limit of $T_{n}\langle 
\mathcal{E}_{n}\rangle _{n}$ for $\varepsilon \rightarrow 0$ at fixed $%
\Lambda $ we get zero, because by assumption (II) $\langle \mathcal{E}%
_{n}\rangle _{n}$ tends to zero for $\varepsilon \rightarrow 0$ at fixed $%
\Lambda $. We conclude that the local functional $\mathcal{E}_{n+1}\equiv
T_{n}\mathcal{E}_{n}$ is truly $\varepsilon $ evanescent at fixed $\Lambda $%
. Therefore, assumption (II) is also promoted to $n+1$ loops.

Since all inductive assumptions have been successfully promoted to $n+1$
loops, the DHD-renormalized action $S_{R}=S_{\infty }$ satisfies 
\begin{equation*}
(S_{R},S_{R})=\mathcal{E}_{R}\text{,}
\end{equation*}%
where $\langle \mathcal{E}_{R}\rangle $ vanishes in the DHD\ limit, because
it contains only the terms of (\ref{survo}) except $\varepsilon ^{0}\Lambda
^{0}$ and $\varepsilon ^{0}/\Lambda $. Finally, the DHD-renormalized $\Gamma 
$ functional $\Gamma _{R}=\Gamma _{\infty }$ is such that the anomaly
functional 
\begin{equation*}
\mathcal{A}_{R}=(\Gamma _{R},\Gamma _{R})=\left\langle
(S_{R},S_{R})\right\rangle _{S_{R}}=\langle \mathcal{E}_{R}\rangle =\mathcal{%
O}(\varepsilon )
\end{equation*}%
tends to zero in the DHD limit, which means that gauge anomalies cancel out
to all orders.

The DHD framework defines a subtraction scheme where the cancellation takes
place naturally and manifestly. In any other framework, the right scheme
must be identified step-by-step, from two loops onwards, by fine-tuning
local counterterms.

Some final comments are in order. Because of (\ref{rena}) higher-order
divergent terms of the form $\Lambda ^{p}$ln$^{k}\Lambda /\varepsilon $ are
generated along the way. They appear in $S_{R}$ and in the partially
renormalized actions $S_{n}$. Our renormalization procedure (which is just
made of redefinitions of parameters, fields and sources) makes them cancel
opposite contributions coming from diagrams. Therefore, they do not appear
in the $\Gamma $ functionals $\Gamma _{R}$ and $\Gamma _{n}$, which are
indeed regular in the limit $\varepsilon \rightarrow 0$ at $\Lambda $ fixed.

In several steps of the proof we have used the fact that $S_{\Lambda }=S_{r}+%
\mathcal{O}(1/\Lambda ^{6})$. It is important that the higher-derivative
regularized classical action $S_{\Lambda }$ does not contain terms with
fewer inverse powers of $\Lambda $. Consistently with this, renormalization
does not require to turn them on. The operations $T_{n}$ may contain
powerlike divergences, which can generate terms with less than 6 inverse
powers of $\Lambda $ when they act on $S_{\Lambda }-S_{r}$. Those terms are
at least one loop and not divergent, so they do not affect the structure of
the classical action $S_{\Lambda }$.

\section{Standard Model and more general theories}

\setcounter{equation}{0}

In this section we show how to extend the proof of the previous sections to
the standard model and the most general perturbatively unitary,
power-counting renormalizable theories. We just need to include photons $%
V_{\mu }$, scalar fields $\varphi $ and right-handed fermions $\chi _{R}$,
which were dropped so far for simplicity. Depending on the representations,
we can also add Majorana masses to the fermions $\psi _{L}$.

We begin from the fermions. The starting classical action (\ref{elle}) is
modified as follows: 
\begin{equation*}
S_{c}\rightarrow S_{c}+\int \bar{\chi}_{R}^{I}\imath \slashed{D}\chi
_{R}^{I}+S_{m},
\end{equation*}%
where $S_{m}$ collects the mass terms, when allowed by the representations: 
\begin{equation}
S_{m}=-\int \left( \bar{\chi}_{R}^{I}m_{IJ}\psi _{L}^{J}+\bar{\psi}%
_{L}^{I}m_{JI}^{\ast }\chi _{R}^{J}\right) -\int (\bar{\psi}%
_{L}^{cI}M_{IJ}\psi _{L}^{J}+\bar{\psi}_{L}^{I}M_{JI}^{\ast }\psi
_{L}^{cJ})-\int (\bar{\chi}_{R}^{cI}M_{IJ}^{\prime }\chi _{R}^{J}+\bar{\chi}%
_{R}^{I}M_{JI}^{\ast \prime }\chi _{R}^{cJ}).  \label{massa}
\end{equation}%
The functional $S_{K}$ that collects the symmetry transformations is also
extended: 
\begin{equation*}
S_{K}\rightarrow S_{K}+g\int \left( \bar{\chi}_{R}^{I}T^{a}C^{a}K_{\chi
}^{I}+\bar{K}_{\chi }^{I}T^{a}C^{a}\chi _{R}^{I}\right) .
\end{equation*}%
Clearly, $\Psi $ and $(S_{K},\Psi )$ are unmodified. To regularize the
right-handed fermions we mirror what we did for the left-handed ones. In the
same way as we added partners $\psi _{R}$ for $\psi _{L}$ that decouple in
four dimensions, we add partners $\chi _{L}$ for $\chi _{R}$ that also
decouple when $D\rightarrow 4$. The correction to $S_{\text{LR}}$ is 
\begin{equation*}
S_{\text{LR}}\rightarrow S_{\text{LR}}+\varsigma _{IJ}^{\prime }\int \bar{%
\chi}_{L}^{I}\imath \slashed{\partial}\chi _{R}^{J}+\varsigma _{JI}^{\prime
\ast }\int \bar{\chi}_{R}^{I}\imath \slashed{\partial}\chi _{L}^{J}+\int 
\bar{\chi}_{L}^{I}\imath \slashed{\partial}\chi _{L}^{I}.
\end{equation*}%
Massive terms involving the regularizing partners $\psi _{R}^{I}$ and $\chi
_{L}^{I}$ can also be added. Differently from (\ref{massa}), they are not
renormalized, so their coefficients must be independent of the ones
appearing in (\ref{massa}). The evanescent corrections $S_{\text{ev}}$ of
formula (\ref{sev}) are affected only in the sector $S_{c\text{ev}}$, which
is extended to include terms such as the integrals of 
\begin{equation}
\bar{\chi}_{R}^{I}\imath \slashed{\partial}\psi _{L}^{J},\qquad \bar{\psi}%
_{L}^{I}\imath \slashed{\partial}\chi _{R}^{J},\qquad \bar{\chi}_{R}^{I}%
\slashed{A}\psi _{L}^{J},\qquad \bar{\psi}_{L}^{I}\slashed{A}\chi _{R}^{J},
\label{scev1}
\end{equation}%
multiplied by independent constants. Evanescent terms of the Majorana type
may also be allowed.

Next, we add the higher-derivative regularizing terms 
\begin{equation*}
\int \bar{\chi}_{R}^{I}\imath \slashed{D}\left[ \left( \frac{D^{2}}{\Lambda
^{2}}\right) ^{\!\!3}-\left( \frac{\partial ^{2}}{\Lambda ^{2}}\right)
^{\!\!3}\right] \chi _{R}^{I}+\int \bar{\chi}^{I}\imath \slashed{\partial}%
\left( \frac{\partial ^{2}}{\Lambda ^{2}}\right) ^{\!\!3}\chi ^{I}
\end{equation*}%
to $S_{\Lambda }$, where $\chi ^{I}=\chi _{L}^{I}+\chi _{R}^{I}$. The gauge
fermion $\Psi _{\Lambda }$ does not change, as well as $S_{\Lambda \hspace{%
0.01in}\text{ev}}-S_{c\text{ev}}$. Tilde fields and sources are defined as
before and every argument of the proof can be extended straightforwardly.
Now, wave-function renormalization constants can mix right-handed fermions
with conjugates of left-handed ones. The contributions of right-handed
fermions to the one-loop anomalies $\mathcal{A}_{f\hspace{0.01in}\text{nev}%
}^{(1)}=\mathcal{A}_{\Lambda \hspace{0.01in}\text{nev}}^{(1)}$ are given by
a formula similar to (\ref{a1loop}), the only difference being that the
trace appearing in the Bardeen term on the right-hand side is calculated on
the appropriate representations $T_{R}^{a}$ ($C\rightarrow C^{a}T_{R}^{a}$, $%
A_{\mu }\rightarrow A_{\mu }^{a}T_{R}^{a}$) and is multiplied by a further
minus sign. The one-loop gauge anomalies $\mathcal{A}_{f\hspace{0.01in}\text{%
nev}}^{(1)}$ are trivial when the Bardeen terms cancel out in the total, and
there exists a local functional $\chi (gA)$ such that $\mathcal{A}_{f\hspace{%
0.01in}\text{nev}}^{(1)}=(S_{K},\chi )$.

Scalars can be added\ by making the replacements 
\begin{eqnarray*}
S_{c} &\rightarrow &S_{c}+\int (D_{\mu }\varphi )^{\dagger }(D^{\mu }\varphi
)+m^{2}\int \varphi ^{\dagger }\varphi +\frac{\lambda }{4}\int (\varphi
^{\dagger }\varphi )^{2}+S_{Y}, \\
S_{K} &\rightarrow &S_{K}-g\int \left( \varphi ^{\dagger
}T^{a}C^{a}K_{\varphi }+K_{\varphi }^{\dagger }T^{a}C^{a}\varphi \right) ,
\end{eqnarray*}%
where $S_{Y}$ denotes the Yukawa terms. As before, the renormalized action
is linear in the sources $K$, by ghost number conservation and power
counting. The evanescent corrections $S_{c\text{ev}}$ include new terms such
as the integrals of 
\begin{equation}
(\partial _{\hat{\mu}}\varphi )^{\dagger }(\partial ^{\hat{\mu}}\varphi
),\qquad (\partial _{\hat{\mu}}\varphi )^{\dagger }T^{a}(A^{\hat{\mu}%
a}\varphi ),  \label{scev2}
\end{equation}%
while $S_{\Lambda \hspace{0.01in}\text{ev}}-S_{c\text{ev}}$ does not change.
The higher-derivative regularizing terms are 
\begin{equation*}
\int (D_{\mu }\varphi )^{\dagger }\left( \frac{D^{2}}{\Lambda ^{2}}\right)
^{4}(D^{\mu }\varphi ),
\end{equation*}%
so the tilde fields and sources 
\begin{equation*}
\tilde{\varphi}=\frac{\varphi }{\Lambda ^{4}},\qquad \tilde{K}_{\varphi
}=\Lambda ^{4}K_{\varphi },
\end{equation*}%
are such that $[\tilde{g}\tilde{\varphi}]=5$, $[\tilde{g}\tilde{K}_{\varphi
}]=14$. With these choices, the matter fields and their sources still cannot
contribute to the one-loop counterterms $\tilde{\Gamma}_{\Lambda \hspace{%
0.01in}\text{div}}^{(1)}$ of the higher-derivative theory $\tilde{S}%
_{\Lambda }$, nor to the nonevanescent one-loop gauge anomalies $\mathcal{%
\tilde{A}}_{\Lambda \hspace{0.01in}\text{nev}}^{(1)}$. Moreover, we still
have $S_{\Lambda }-S_{r}=\mathcal{O}(1/\Lambda ^{6})$. Therefore, all
arguments used in the proof of the previous sections generalize
straightforwardly.

Finally, we add photons. Assume that the group $G$ contains $N$ $U(1)$
factors and denote their gauge fields with $V_{\mu }^{u}$, $u=1,\ldots N$.
Then make the replacements 
\begin{eqnarray*}
S_{c} &\rightarrow &S_{c}-\frac{1}{4}\int \zeta _{uv}W_{\mu \nu }^{u}W^{v\mu
\nu },\qquad D_{\mu }\pi ^{I}\rightarrow D_{\mu }\pi ^{I}+iQ^{u}V_{\mu
}^{u}\pi ^{I}, \\
S_{K} &\rightarrow &S_{K}-\int (\partial _{\mu }C^{u})K^{\mu u}-\imath g\int
C^{u}\sum_{\pi }(\pi ^{I\dagger }QK_{\pi }^{I}-K_{\pi }^{I\dagger }Q\pi
^{I}),
\end{eqnarray*}%
where $W_{\mu \nu }^{u}=\partial _{\mu }V_{\nu }^{u}-\partial _{\nu }V_{\mu
}^{u}$, $\zeta _{uv}$ is an invertible constant matrix, $\pi ^{I}$ is any
matter field in the irreducible representation $R^{I}$ of $G$, and $\pi
^{I\dagger },K_{\pi }^{I\dagger }$ stand for $\bar{\pi}^{I},\bar{K}_{\pi
}^{I}$ if $\pi ^{I}$ is a fermion. We define extended $G$ indices $\hat{a},%
\hat{b},\ldots $ to include both sets of indices $u,v,\ldots $ and $%
a,b,\ldots $, and write $A_{\mu }^{\hat{a}}=\{V_{\mu }^{u},A_{\mu }^{a}\}$.
The $U(1)$ charges of matter fields are denoted by $gq_{I}^{u}$. We also
write $T^{\hat{a}}=\{iQ^{u},T^{a}\}$, where $Q^{u}$ acts on $\pi ^{I}$ by
multiplying it by $q_{I}^{u}$. The change of the gauge fermion (\ref{gf})\
is 
\begin{equation*}
\Psi (\Phi )\rightarrow \Psi (\Phi )+\int \bar{C}^{u}\left( \partial ^{\mu
}A_{\mu }^{u}+\frac{\xi _{uv}}{2}B^{v}\right) .
\end{equation*}%
The sector $S_{c\hspace{0.01in}\text{ev}}$ of $S_{\hspace{0.01in}\text{ev}}$
is also extended, to include $V$-dependent evanescent terms similar to those
already met in (\ref{scev}), (\ref{scev1}) and (\ref{scev2}). Instead, $%
S_{\Lambda \hspace{0.01in}\text{ev}}-S_{c\text{ev}}$ remains the same, since
the $U(1)$ ghosts decouple.

The action $S_{c\Lambda }$ is extended to include the higher-derivative
regularizing terms 
\begin{equation*}
-\frac{1}{4}\int W_{\mu \nu }^{u}\left( \frac{\partial ^{2}}{\Lambda ^{2}}%
\right) ^{\!\!8}W^{u\hspace{0.01in}\mu \nu },
\end{equation*}%
while the change of gauge fermion is 
\begin{equation*}
\Psi _{\Lambda }(\Phi )\rightarrow \Psi _{\Lambda }(\Phi )+\int \bar{C}%
^{u}\left( Q\left( \Box \right) \partial ^{\mu }V_{\mu }^{u}+\frac{1}{2}%
P_{uv}\left( \Box \right) B^{v}\right) ,
\end{equation*}%
where 
\begin{equation*}
P_{uv}\left( \Box \right) =\xi _{uv}+\frac{\delta _{uv}\xi ^{\prime }}{%
\Lambda ^{16}}\Box ^{8}.
\end{equation*}%
Finally, $S_{\Lambda \hspace{0.01in}\text{ev}}$ inherits the modifications
made on $S_{c\text{ev}}$. Tilde fields and sources are defined as before.
The one-loop renormalization of the higher-derivative theory $\tilde{S}%
_{\Lambda }$ is made of the replacements (\ref{rena}) plus similar
replacements 
\begin{equation*}
\zeta _{uv}\rightarrow \zeta _{uv}+\frac{f_{uv}}{\varepsilon }g^{2}
\end{equation*}%
for $\zeta _{uv}$, where $f_{uv}$ are calculable constants.

Let us describe the nontrivial contributions to the one-loop gauge anomalies 
$\mathcal{A}_{f\hspace{0.01in}\text{nev}}^{(1)}$. We have terms of the
Badreen type and terms proportional to $C^{u}W_{\mu \nu }^{v}W^{z\mu \nu }$.
Using differential forms, the terms of the Bardeen type are linear
combinations of $\mathcal{B}_{1}=\int \mathrm{Tr}\left[ dC{\scriptscriptstyle%
\wedge }A{\scriptscriptstyle\wedge }dA\right] $ and $\mathcal{B}_{2}=\int 
\mathrm{Tr}\left[ dC{\scriptscriptstyle\wedge }A{\scriptscriptstyle\wedge }A{%
\scriptscriptstyle\wedge }A\right] $, as in formula (\ref{a1loop}), where
now $C=C^{\hat{a}}T_{f}^{\hat{a}}$, $A=A_{\mu }^{\hat{a}}T_{f}^{\hat{a}%
}dx^{\mu }$, $d=dx^{\mu }\partial _{\mu }$ and $T_{f}^{\hat{a}}$ are the
matrices $T^{\hat{a}}$ restricted to the fermions. The coefficient of $%
\mathcal{B}_{1}$ is the same as in formula (\ref{a1loop}), apart from the
minus sign associated with right-handed fermions. The coefficient of $%
\mathcal{B}_{2}$ is uniquely determined by the coefficient of $\mathcal{B}%
_{1}$, but it differs from the one of formula (\ref{a1loop}) any time $U(1)$
gauge fields and/or ghosts are involved. The terms proportional to $CW_{\mu
\nu }W^{\mu \nu }$ can only appear in (unusual) situations where global $%
U(1) $ gauge symmetries are potentially anomalous. One-loop gauge anomalies
are trivial when all these terms cancel out, and there exists a\ local
functional $\chi (gA)$ such that $\mathcal{A}_{f\hspace{0.01in}\text{nev}%
}^{(1)}=(S_{K},\chi )$.

The correction to the canonical transformation (\ref{op1}) reads 
\begin{equation*}
F_{n}(\Phi ,K^{\prime })\rightarrow F_{n}(\Phi ,K^{\prime })+\int (V_{\mu
}^{u}Z_{n\hspace{0.01in}uv}^{1/2}K^{\prime \hspace{0.01in}\mu v}+C^{u}Z_{n%
\hspace{0.01in}uv}^{1/2}K_{C}^{\prime \hspace{0.01in}v}+\bar{C}^{u}Z_{n%
\hspace{0.01in}uv}^{-1/2}K_{\bar{C}}^{\prime \hspace{0.01in}v}+B^{u}Z_{n%
\hspace{0.01in}uv}^{-1/2}K_{B}^{\prime \hspace{0.01in}v}),
\end{equation*}%
and the redefinitions (\ref{op2}) are accompanied by 
\begin{equation*}
q_{I}^{u\hspace{0.01in}\prime }=Z_{n\hspace{0.01in}uv}^{-1/2}q_{I}^{v},%
\qquad \xi _{uv}^{\prime }=Z_{n\hspace{0.01in}uw}^{1/2}\xi _{wz}Z_{n\hspace{%
0.01in}zv}^{1/2},
\end{equation*}%
so that the $U(1)$ gauge-fixing sector $(S_{K},\Psi )$, including the ghost
action, as well as the $U(1)$ sector of $S_{K}$, are nonrenormalized.

With the rules of this section gauge anomalies manifestly cancel to all
orders in the most general perturbatively unitary, renormalizable gauge
theory coupled to matter, as long as they vanish at one loop. We stress
again that the proof we have given also works when the theory is conformal
or finite, or the first coefficients of its beta functions vanish, where
instead RG techniques are powerless.

\section{Conclusions}

\setcounter{equation}{0}

We have reconsidered the Adler-Bardeen theorem, focusing on the cancellation
of gauge anomalies to all orders, when they are trivial at one loop. The
proof we have worked out is more powerful than the ones appeared so far and
makes us understand aspects that the previous derivations were unable to
clarify. Key ingredients of our approach are the Batalin-Vilkovisky
formalism and a regularization technique that combines the dimensional
regularization with the higher-derivative gauge invariant regularization.
The most important result is the identification of the subtraction scheme
where gauge anomalies manifestly cancel to all orders. We have not used
renormalization-group arguments, so our results apply to the most general
perturbatively unitary, renormalizable gauge theories coupled to matter,
including conformal field theories, finite theories, and theories where the
first coefficients of the beta functions vanish.

In view of future generalizations to wider classes of quantum field
theories, we have paid attention to a considerable amount of details and
delicate steps that emerge along with the proof. We are convinced that the
techniques developed here may help us identify the right tools to upgrade
the formulation of quantum field theory and simplify the proofs of all-order
theorems.

\renewcommand{\thesection}{A}

\section*{Appendix A. Calculation of one-loop anomalies}
\setcounter{equation}{0}
\renewcommand{\theequation}{\thesection.\arabic{equation}}

In this appendix we illustrate our approach by calculating the one-loop
coefficient of the Bardeen anomaly in chiral gauge theories. That
coefficient is scheme independent, so we can work at $\Lambda =\infty $,
which means use the dimensionally regularized action $S_{r}$ of (\ref{exta}%
). Actually, we can equivalently use the action $S_{r0}$ of (\ref{sr}),
because it is easy to check that the contributions due to $S_{\text{ev}}$ do
not contain fermion loops. Therefore, they cannot generate the tensor $%
\varepsilon ^{\mu \nu \rho \sigma }$.

For simplicity, we first work with chiral QED and then generalize the result
to non-Abelian theories. The action\ reads 
\begin{eqnarray}
S_{r0}(\Phi ,K) &=&-\frac{1}{4}\int F_{\mu \nu }F^{\mu \nu }+\int \bar{\psi}%
\imath \gamma ^{\mu }\partial _{\mu }\psi -\int q_{L}\bar{\psi}_{L}\gamma
^{\mu }A_{\mu }\psi _{L}+(S_{K},\Psi )+S_{K},  \notag \\
S_{K} &=&-\int (\partial _{\mu }C)K^{\mu }+\imath q_{L}\int (\bar{\psi}%
_{L}CK_{\psi }+\bar{K}_{\psi }C\psi _{L})-\int BK_{\bar{C}},  \label{schir}
\end{eqnarray}%
where $q_{L}$ is the charge and the gauge fermion is 
\begin{equation*}
\Psi =\int \bar{C}\left( \partial ^{\mu }A_{\mu }+\frac{\xi }{2}B\right) .
\end{equation*}%
We have 
\begin{eqnarray*}
(S_{r0},S_{r0}) &=&-2q_{L}\int C\left( \bar{\psi}_{L}\hat{\slashed{\partial}}%
\psi _{R}+(\partial _{\hat{\mu}}\bar{\psi}_{R})\gamma ^{\hat{\mu}}\psi
_{L}\right) \\
&=&2\int C(\partial _{\mu }J^{\mu })+2\imath q_{L}\int C\left( \bar{\psi}_{L}%
\frac{\delta _{l}\bar{S}}{\delta \bar{\psi}_{L}}-\frac{\delta _{r}\bar{S}}{%
\delta \psi _{L}}\psi _{L}\right) ,
\end{eqnarray*}%
where $J^{\mu }=q_{L}\bar{\psi}_{L}\gamma ^{\mu }\psi _{L}$ is the gauge
current and $\bar{S}(\Phi )=S(\Phi ,0)$.

We focus on the matter-independent contributions $\mathcal{A}_{B}$ to the
anomaly $\mathcal{A}=\langle (S_{r0},S_{r0})\rangle _{S_{r0}}$, so we can
take the ghosts outside the average. Switching to momentum space, we get 
\begin{equation*}
\mathcal{A}_{B}=-2\imath q_{L}\int \frac{\mathrm{d}^{D}p}{(2\pi )^{D}}C(-k)\,%
\text{tr}\!{\left[ \hat{\slashed{p}}(P_{R}-P_{L})\langle \psi (p+k_{1})\bar{%
\psi}(-p+k_{2})\rangle \right] .}
\end{equation*}%
Here and below the integrals on momenta $k$ in $\mathcal{A}_{B}$ are
understood. We expand the fermion two-point function in powers of the gauge
field. The linear term gives a contribution that by power counting and ghost
number conservation is proportional to 
\begin{equation*}
\int (\Box C)(\partial ^{\mu }A_{\mu })=(S_{K},\chi ^{\prime }),\qquad \chi
^{\prime }=\frac{1}{2}\int (\partial _{\mu }A_{\nu })(\partial ^{\mu }A^{\nu
}).
\end{equation*}%
It can be subtracted away as explained in formula (\ref{suba}). Then we
concentrate on the contributions $\mathcal{A}_{B}^{\prime }$ to $\mathcal{A}%
_{B}$ that are quadratic in the gauge field. We observe that one fermion
propagator is sandwiched between two $P_{L}$'s or two $P_{R}$'s, which
projects its numerator onto the evanescent sector, and the other two
propagators are sandwiched between $P_{L}$ and $P_{R}$, which projects their
numerators onto the physical sector. We get 
\begin{equation*}
\mathcal{A}_{B}^{\prime }=-2q_{L}^{3}\int \frac{\mathrm{d}^{D}p}{(2\pi )^{D}}%
\frac{C(-k)\hat{p}^{2}}{(p+k_{1})^{2}p^{2}(p-k_{2})^{2}}\hspace{0.01in}\text{%
tr}\!{\left[ P_{L}\slashed{k}\slashed{A}(k_{1})\overline{\slashed{p}}%
\slashed{A}(k_{2})\right] .}
\end{equation*}%
The photons and their momenta $k_{1}$, $k_{2}$ can be taken to be strictly
four dimensional. Turning to Euclidean space and using 
\begin{equation*}
I^{\mu }=\int_{Eucl}\frac{\mathrm{d}^{D}p}{(2\pi )^{D}}\frac{\hat{p}^{2}\bar{%
p}^{\mu }}{(p+k_{1})^{2}p^{2}(p-k_{2})^{2}}=\frac{1}{96\pi ^{2}}(k_{1}^{\mu
}-k_{2}^{\mu })+\mathcal{O}(\varepsilon )\text{,}
\end{equation*}%
we obtain 
\begin{equation*}
\mathcal{A}_{B}^{\prime }=-\frac{q_{L}^{3}}{12\pi ^{2}}\int C(-k)\varepsilon
^{\mu \nu \rho \sigma }k_{1\mu }A_{\nu }(k_{1})k_{2\rho }A_{\sigma }(k_{2}),
\end{equation*}%
where $\varepsilon ^{0123}=1$. Converting to coordinate space and including
the trivial contributions, we finally get 
\begin{equation*}
\mathcal{A}_{B}=\frac{q_{L}^{3}}{48\pi ^{2}}\int C\varepsilon ^{\mu \nu \rho
\sigma }F_{\mu \nu }F_{\rho \sigma }+(S_{K},\chi ).
\end{equation*}%
After subtraction of the trivial terms the divergence of the current
averages to 
\begin{equation*}
\langle \partial _{\mu }J^{\mu }\rangle =\frac{q_{L}^{3}}{96\pi ^{2}}%
\varepsilon ^{\mu \nu \rho \sigma }F_{\mu \nu }F_{\rho \sigma }-\imath
q_{L}\left( \bar{\psi}_{L}\frac{\delta _{l}\bar{S}}{\delta \bar{\psi}_{L}}-%
\frac{\delta _{r}\bar{S}}{\delta \psi _{L}}\psi _{L}\right) .
\end{equation*}%
Incidentally, the calculation shows that $\mathcal{A}_{B}$ receives no
contributions proportional to $\int CF^{\mu \nu }F_{\mu \nu }$. This term is
in principle allowed by the cohomological constraint (\ref{aqua}) in Abelian
theories, but actually does not show up. If it did, it would imply that the
global symmetry associated with the gauge symmetry is anomalous, which is of
course not true.

The calculation just done also proves formula (\ref{a1loop}), after
inserting matrices $T^{a}$ and structure constants $f^{abc}$ where
appropriate.

\renewcommand{\thesection}{B}

\section*{Appendix B. Formula of the anomaly functional}
\setcounter{equation}{0}
\renewcommand{\theequation}{\thesection.\arabic{equation}}

In this appendix we recall the proof of the last equalities of formulas (\ref%
{anom}) and (\ref{an}), which express the anomaly functional $\mathcal{A}$.
We show that%
\begin{equation}
(\Gamma ,\Gamma )=\langle (S,S)\rangle \hspace{0.01in},  \label{anomapp}
\end{equation}%
where $S$ is a dimensionally regularized action, the average is defined by
the functional integral (\ref{zg}) and $\Gamma $ is the Legendre transform
of $W$, defined by the same integral.

Recall that, using the dimensional regularization technique, local
perturbative field redefinitions have Jacobian determinants identically
equal to one. Indeed, from the diagrammatic point of view such Jacobians are
equal to 1 plus integrals of polynomials of the momenta $p$ in $\mathrm{d}%
^{D}p$, which vanish. Now, if we make the change of field variables%
\begin{equation*}
\Phi ^{\alpha }\rightarrow \Phi ^{\alpha }+\theta (S,\Phi ^{\alpha })=\Phi
^{\alpha }-\theta \frac{\delta _{r}S}{\delta K_{\alpha }}
\end{equation*}%
in the functional integral (\ref{zg}), where $\theta $ is a constant
anticommuting parameter, we obtain 
\begin{equation*}
-\iota \theta \int \left\langle \frac{\delta _{r}S}{\delta K_{\alpha }}\frac{%
\delta _{l}S}{\delta \Phi ^{\alpha }}\right\rangle -\iota \theta \int
\left\langle \frac{\delta _{r}S}{\delta K_{\alpha }}\right\rangle J_{\alpha
}=0.
\end{equation*}%
Using this identity, and recalling that the two terms of the antiparentheses
(\ref{usa}) are equal when $X$ and $Y$ coincide and have bosonic statistics,
we get 
\begin{equation*}
\frac{1}{2}\langle (S,S)\rangle =-\int \left\langle \frac{\delta _{r}S}{%
\delta K_{\alpha }}\frac{\delta _{l}S}{\delta \Phi ^{\alpha }}\right\rangle
=\int \left\langle \frac{\delta _{r}S}{\delta K_{\alpha }}\right\rangle
J_{\alpha }.
\end{equation*}%
The average of $\delta _{r}S/\delta K_{\alpha }$ is equal to $\delta
_{r}W/\delta K_{\alpha }$, which is also $\delta _{r}\Gamma /\delta
K_{\alpha }$, because the sources $K$ are inert in the Legendre transform
that defines $\Gamma $. Using $J_{\alpha }=-\delta _{l}\Gamma /\delta \Phi
^{\alpha }$, we arrive at 
\begin{equation*}
\frac{1}{2}\langle (S,S)\rangle =\int \frac{\delta _{r}W}{\delta K_{\alpha }}%
J_{\alpha }=-\int \frac{\delta _{r}\Gamma }{\delta K_{\alpha }}\frac{\delta
_{l}\Gamma }{\delta \Phi ^{\alpha }}=\frac{1}{2}(\Gamma ,\Gamma ).
\end{equation*}

For other details, see for example the appendix of ref. \cite{back1}. Note
that the dimensional regularization is crucial for the derivation. Clearly,
formula (\ref{anomapp}) also works if we use the DHD regularization, because
the dimensional one is embedded in it. We then obtain formula (\ref{an}).

If a dimensionally regularized action $S$ satisfies $(S,S)=0$ in arbitrary $%
D=4-\varepsilon$ dimensions, then gauge anomalies are manifestly absent, as
in QED and QCD, and formula (\ref{anom}) correctly gives $\mathcal{A}=0$.
When chiral fermions are present, as in the standard model, we have the $%
\gamma_{5}$ problem. A dimensionally regularized action $S$ cannot equip
chiral fermions with well-behaved propagators, and satisfy $(S,S)=0$ in $D$
dimensions at the same time. The na\"{\i}ve fermionic propagators, given by
the starting action (\ref{elle}), do not depend on the evanescent components 
$\hat{p}$ of momenta. Then, according to the rules of the dimensional
regularization, fermion loops integrate to zero, which means that the the
starting action (\ref{elle}) is not well regularized. The action must be
modified to equip fermions with well-behaved propagators, for example by
adding the correction $S_{\text{LR}}$ of formula (\ref{slr}). Once this is
done, however, $S$ satisfies $(S,S)=\mathcal{O}(\varepsilon )$, as shown in
formulas (\ref{srsr}) and (\ref{srsr2}). The evanescent terms $\mathcal{O}%
(\varepsilon )$, inserted in the diagrams belonging to the average $%
\left\langle (S,S)\right\rangle =\left\langle \mathcal{O}(\varepsilon
)\right\rangle $, can simplify poles $1/\varepsilon $ and give finite,
potentially anomalous contributions, as shown in the calculation of the
previous appendix.

It is worth to stress that our investigation only concerns gauge anomalies,
so $\mathcal{A}=0$ does not exclude the presence of other types of
anomalies, such as the axial anomaly of QED.

\end{document}